\documentclass[11pt]{article}
\usepackage{amsmath}
\usepackage{amssymb}
\usepackage{latexsym}
\input feynman
\input prepictex
\input pictexwd
\input postpictex
\numberwithin{equation}{section}

\def\D{\Delta}
\def\L{\Lambda}
\def\S{\Sigma}
\def\a{\alpha}

\def\d{\delta}
\def\e{\varepsilon}
\def\x{{\bf x}}
\def\m{\mu}
\def\n{\nu}
\def\s{\sigma}
\def\r{\rho}
\def\l{\lambda}
\def\t{\tau}
\def\o{\omega}
\def\v{\varrho}
\def\mc{\mathcal}
\def\N{\nabla}
\parskip=\medskipamount

\begin{document}
\renewcommand{\thefootnote}{\fnsymbol{footnote}}
\begin{titlepage}
\thispagestyle{empty}
\vspace*{1.5cm}
\large{
\begin{flushright}
LMU-TPW 00--15\\
hep-th/0007061
\end{flushright}}
\vspace*{5mm}
\begin{center}
{\bf\Large Cubic couplings in $D=6$ ${\mc N}=4b$ supergravity
on $AdS_3\times S^3$}
\\
\vspace*{0.5cm}
G. Arutyunov$^a$
\footnote[1]{On leave of absence from Steklov Mathematical Institute,
Gubkin str.8, GSP-1, 117966, Moscow, Russia}
\footnote[2]{arut@theorie.physik.uni-muenchen.de},
A. Pankiewicz$^a$
\footnote[3]{ari@theorie.physik.uni-muenchen.de}
\ and\ S. Theisen$^a$\footnote[4]{theisen@theorie.physik.uni-muenchen.de}
\\
\vspace*{0.4cm}
{\small $^a$ Sektion Physik,\\
Universit\"at M\"unchen, Theresienstr. 37,\\
D-80333 M\"unchen, Germany}
\end{center}
\vspace*{1cm}

\begin{abstract}
We determine the AdS exchange diagrams needed for the computation of
4--point functions of chiral primary operators in the SCFT$_2$
dual to the $D=6$ ${\mc N}=4b$ supergravity on the $AdS_3\times S^3$ 
background and compute the corresponding cubic couplings. 
We also address the issue of consistent truncation.
\end{abstract}
\end{titlepage}
\renewcommand{\thefootnote}{\arabic{footnote}}
\section{Introduction}
The AdS/CFT correspondence provides information
about the strong coupling behaviour of some conformal field
theories by studying their supergravity (string) duals \cite{mal}--\cite{aha}.
In particular, the AdS/CFT duality relates type IIB string theory on 
$AdS_3\times S^3\times M^4$, where $M^4$
is either $K3$ or $T^4$, to a certain ${\mc N}=(4,4)$
supersymmetric two--dimensional
conformal field theory (CFT) living on the boundary of $AdS_3$.
A two--dimensional sigma model with the target space being a deformation of
the orbifold symmetric product $S^N(M^4)=(M^4)^N/S_N$, $N\to \infty$,
is believed to provide an effective description of this CFT \cite{doug}.

An important class of operators in the supersymmetric
CFT are the Chiral Primary
Operators (CPOs) since they are annihilated by
$1/2$ of the supercharges and in two dimensions their highest weight
components of the R--symmetry group form a ring.
On the gravity side the CPOs correspond to Kaluza--Klein (KK) modes of the
type IIB supergravity compactification.
Recently, using the orbifold technique developed in \cite{gs}
the three--point functions of scalar CPOs were computed \cite{jev} in the
CFT on the symmetric product $S^N(M^4)$
and, on the other hand, in  $AdS_3\times S^3$ supergravity \cite{mih},
and were found
to disagree\footnote{
Computing the 2-- and  3--point functions of CPOs in the supergravity
approximation by using the prescription \cite{gub}, which is known
to be compatible with the Ward identities \cite{fr},
one finds a result different from \cite{mih}.
This however does not remove the disagreement between
CFT and gravity calculations.}. On the other hand, computations of
quantities that are stable under deformations of the orbifold CFT,
like the spectrum
of the CPOs and the elliptic genus, were found to be in
complete agreement \cite{b}.
Obviously this supports
the expectation that the $AdS_3\times S^3$ background
may correspond to some deformation of the target space
$S^N(M^4)$ of the boundary CFT.
However, even though one presently does not have an explicit
sigma model formulation of the
boundary CFT (see also \cite{lm} for recent developements), one may proceed to study the CFT by using
directly the gravity dual description
and the AdS/CFT correspondence \cite{kut}.

In this paper we study the $AdS_3\times S^3$ compactification of the
$D=6$ ${\mc N}=4b$ supergravity coupled to $n$ tensor multiplets.
In particular, the case $n=21$ corresponds to the theory obtained by
dimensional reduction of type IIB supergravity on $K3$.
Our final aim will be to find the 4--point correlation
functions of the scalar CPOs in
the supergravity approximation. This program was successfully
performed for the ${\mc N}=4$
SYM$_4$ which is related to the $AdS_5\times S^5$
compactification of type IIB supergravity
and led to an understanding of the structure of the
Operator Product Expansion in the field theory
at strong coupling \cite{fr}, \cite{sei}--\cite{liu1}.
As a first necessary step in this direction we derive the
effective gravity action on
$AdS_3$ that contains all cubic couplings involving at
least two gravity fields corresponding
to CPOs in the boundary CFT.

Since the supergravity we consider is a chiral theory it
suffers from the absence of a
simple Lagrangian formulation. In principle, one may approach
the problem of computing
correlation functions by using the Pasti--Sorokin--Tonin
formulation of the six--dimensional
supergravity action, where the manifest Lorentz covariance
is achieved by introducing an
auxilliary scalar field $a$ \cite{tonin}. However, to
obtain the action for physical fields
one needs to fix the gauge symmetries, in particular
the additional symmetry associated with
the field $a$. This breaks the manifest Lorentz covariance
and makes the problem of solving
the noncovariant constraints imposed by gauge fixing
unfeasible. Thus, we prefer to start with
the covariant equations of motion of chiral six--dimensional
supergravity \cite{ro} and
obtain the quadratic, cubic and so on corrections to
the equations of motion for physical
fields by decomposing the original equations near the
$AdS_3\times S^3$ background and
partially fixing the gauge (diffeomorphism) symmetries.
The equations obtained in this way are in general non--Lagrangian with
higher derivative terms and we perform the nonlinear
field redefinitions to remove higher
derivative terms \cite{sei} and bring
the equations to the Lagrangian form.

The spectrum of the $AdS_3\times S^3$
compactification of the $D=6$ ${\mc N}=4b$ supergravity coupled to $n$ 
tensor multiplets
was found in \cite{deg} and
it is governed by the supergroup $SU(1,1|2)_L\times SU(1,1|2)_R$.
Since we are interested in
the quadratic and ultimately in cubic corrections to the
equations of motion for the gravity
fields we reconsider the derivation of the linearized
equations of motion and recover the
spectrum of \cite{deg}. According to \cite{deg}
the scalar CPOs are
divided into two classes. The first class contains CPOs
$\s$ that are singlets with respect
to the internal symmetry group $SO(n)$. The
corresponding gravity fields are mixtures of the trace
of the graviton on $S^3$ and the sphere
components of the self--dual form. The second class
comprise the CPOs $s^r$ transforming in
the fundamental representation of $SO(n)$ and the
corresponding gravity fields are mixtures
of $n$ scalars from the coset space
$SO(5,n)/SO(5)\times SO(n)$ and the sphere components
of the $n$ antiself--dual forms.

As follows from our study, the fields appearing in
the exchange diagrams involving at
least two CPOs may contain, in addition to the CPOs themselves,
also other scalars or vectors, either
in the singlet or in the vector representation of $SO(n)$,
and symmetric 2nd rank (massive)
tensor fields. We find the corresponding cubic couplings. By using the 
factorization property of the Maxwell operator in odd dimensions we
diagonalize the equations
for the vector fields which originate from components of the
second order Einstein equation and
the first order self--duality equation.  This diagonalization
is helpful to identify the
vector fields propagating in the AdS exchange diagrams.
To ensure the wider applicability of our results we
keep $n$ unspecified\footnote{Except $n=21$ another case of interest
is $n=5$. Dimensional reduction of type IIB supergravity on $T^4$
produces the non--chiral $D=6$ ${\mc N}=8$ theory, for which
the equations of motion for the metric,
the scalar fields and the two--forms are the same as for $D=6$ ${\mc N}=4b$
with $n=5$.}.

The cubic couplings exhibit the same vanishing
property in the extremal case
(e.g.\ for three scalar fields $\sigma_k$,
where $k$ denotes a Kaluza--Klein mode, the extremality
condition is $k_1+k_2=k_3$ and permutations thereof)
as the cubic couplings found in the compactification of type IIB supergravity on
$AdS_5\times S^5$ \cite{sei}.

In addition to the cubic couplings involving CPOs we also compute
certain cubic couplings of vector fields, which allows us to
check the consistency of the KK truncation of the
three--dimensional action to the massless graviton multiplet.
Recall that the bosonic part of this multiplet contains the
graviton and the $SU(2)_L\times SU(2)_R$ gauge fields, all of
them carrying non--propagating (topological) degrees of freedom.
Since the other multiplets contain the propagating modes, the
graviton multiplet should admit a consistent truncation and we
show that this is indeed the case. The truncated action coincides
with the topological Chern--Simons action constructed in
\cite{at}. We also consider the problem of the KK truncation to
the sum of two multiplets, one of them the massless graviton
multiplet, whereas the second involves the fields corresponding to
the lowest weight CPOs. Surprisingly, we have found indications that the sum 
of the massless graviton multiplet and the special
spin--$\frac{1}{2}$  multiplet containing the lowest mode scalar CPOs may
admit a consistent truncation. This situation reminds of, but
is different from the $AdS_5\times S^5$ compactification, where the
lowest weight CPOs occur in the stress tensor multiplet, which on
the gravity side corresponds to the massless graviton multiplet,
allowing a consistent truncation \cite{gleb1}. In the $AdS_3$
case the gauge degrees of freedom encoded in the graviton
multiplet give rise to the ${\mc N}=(4,4)$ superconformal algebra
of the boundary CFT \cite{bh}.

The paper is organized as follows. In section 2 we recall
the covariant equations of motion
for the bosonic sector of $D=6$ ${\mc N}=4b$ supergravity \cite{ro},
introduce notation and represent
our results. In section 3 we review the linearized equations
of motion and recover
the spectrum found in \cite{deg}. In section 4 we discuss
the structure of the quadratic
corrections to the linearized equations of motion and explain
the necessary steps
to reduce the equations of motion to a Lagrangian form.
In Appendix A we give the results for
the expansion of the covariant equations of motion up to the second order,
both in spacetime and coset metric perturbations and in Appendix B we establish
the formulae for various
integrals involving spherical harmonics of different kinds.
\section{The cubic effective action in $AdS_3$}
Cubic couplings of chiral primaries may be derived from the
quadratic corrections to the covariant equations of motion
for $D=6$ ${\mc N}=4b$ supergravity coupled to $n$ tensor multiplets \cite{ro}.
All the bosonic fields ---
the graviton, the two--form potentials $B_{MN}^I$, $I=1,\dots ,5+n$
and the scalar sector ---
provide relevant contributions to the quadratic corrections.
The scalar sector constitutes a sigma model over the coset space
$\frac{SO(5,n)}{SO(5)\times SO(n)}$ with vielbein $({V_I}^i,{V_I}^r)$,
$i=1,\dots ,5$,
$r=1,\dots ,n$ which is parameterized by $5n$ scalar fields.
The index $I$ transforms
under global $SO(5,n)$ transformations and is raised and lowered with the
$SO(5,n)$ invariant metric
$\eta=\text{diag}(1_{5\times 5},-1_{n\times n})$, whereas
the indices $(i,r)$ transform
under local composite $SO(5)\times SO(n)$ transformations.
We use the following indices: $M$, $N$ for $D=6$, $\mu$, $\nu$ for
$AdS_3$ and $a$, $b$
for $S^3$ coordinates.

Defining
\begin{equation}\label{cart}
dVV^{-1}=
\begin{pmatrix}
Q^{ij} & \sqrt{2}P^{is} \\
\sqrt{2}P^{rj} & Q^{rs}
\end{pmatrix},
\end{equation}
the covariant derivative in the scalar sector is found
by the Cartan--Maurer equation to be
\begin{equation}
D_MP_N^{ir}=\N_MP_N^{ir}-Q^{ij}_MP_N^{jr}-P_M^{is}Q_N^{sr}
\end{equation}
and the equations of motion for the bosonic sector of $D=6$ ${\mc N}=4b$ 
supergravity are:
\begin{align}
\label{eqn}
R_{MN}&=H^i_{MPQ}{H^i_N}^{PQ}+H^r_{MPQ}{H^r_N}^{PQ}+2P^{ir}_MP^{ir}_N, \\
\label{scalar}
D^MP_M^{ir}&= \frac{\sqrt{2}}{3}H^i_{MPQ}{H^r}^{MP},\\
\label{sat}
*H^i &=H^i,\qquad *H^r=-H^r ,
\end{align}
where
\begin{equation}
\label{fs}
H^i=G^I{V_I}^i,\quad H^r=G^I{V_I}^r\text{ and } G^I=dB^I.
\end{equation}

In units where the radius of $S^3$ is set to unity,
the $AdS_3\times S^3$ background
solution is
\begin{equation}
ds^2=\frac{1}{x_0^2}(dx_0^2+\eta_{ij}dx^idx^j)+d\Omega_3^2,
\end{equation}
where $\eta_{ij}$ is the 2--dimensional Minkowski metric.
One of the self--dual field strengths is singled out and set equal to the
Levi--Cevita tensor, while all others vanish:
\begin{equation}
H^i_{\m\n\r}=\d^{i5}\e_{\m\n\r},\qquad
H^i_{abc}=\d^{i5}\e_{abc},\qquad
H^r_{MNP}=0.
\end{equation}
Here $\e_{\m\n\r}$ and $\e_{abc}$ are the volume forms on
$AdS_3$ and $S^3$, respectively, so that
$\e_{\m\n\r}\e_{abc}$ is the volume
form in six dimensions. The $SO(5,n)$ background vielbein is taken to be constant and by a
global $SO(5,n)$ rotation it can be set to unity.

To construct the Lagrangian equations of motion we represent the fields as
\begin{align}
g_{MN} &= \bar{g}_{MN}+h_{MN}, \\
G^I &= \bar{G}^I + g^I,\quad g^I=db^I, \\ \intertext{and}
{V_I}^i &={\d_I}^i+\phi^{ir}{\d_I}^r+\frac{1}{2}\phi^{ir}\phi^{jr}{\d_I}^j,\\
{V_I}^r &={\d_I}^r+\phi^{ir}{\d_I}^i+\frac{1}{2}\phi^{ir}\phi^{is}{\d_I}^s.
\end{align}

The gauge symmetry of the equations of motion allows one to impose
the de Donder--Lorentz gauge\footnote{From now on all
the covariant derivatives
are understood to be in the background geometry.}:
\begin{equation}\label{donder}
\N^ah_{\m a}=\N^ah_{(ab)}=\N^ab_{Ma}^I=0.
\end{equation}
Here and below the subscript $(ab)$ denotes
symmetrization of indices $a$ and $b$
with the trace removed.

This gauge choice does not fix all the gauge symmetry of the theory,
for a detailed
discussion of the residual symmetry, {\it c.f.} \cite{deg}.
The gauge condition \eqref{donder}
implies that the physical fluctuations are decomposed
in spherical harmonics on $S^3$ as
\footnote{Here and in what follows we use normalized
spherical harmonics,
i.e.
$\int Y^{I_1}Y^{J_1}=\d^{I_1J_1}$,
$\int\bar{g}^{ab}Y_a^{I_3\pm}Y_b^{J_3\pm}=\d^{I_3J_3}$,
$\int\bar{g}^{ac}\bar{g}^{bd}Y_{(ab)}^{I_5\pm}Y_{(cd)}^{J_5\pm}=\d^{I_5J_5}$.}:
\begin{align*}
h_{\m\n}(x,y)&=\sum h_{\m\n}^{I_1}(x)Y^{I_1}(y),
&h_{\m a}(x,y)&=\sum h_{\m}^{I_3\pm}(x)Y^{I_3\pm}_a(y),\\
h^a_a(x,y)&=\sum\pi^{I_1}(x)Y^{I_1}(y),
&h_{(ab)}(x,y)&=\sum\v^{I_5\pm}(x)Y_{(ab)}^{I_5\pm}(y),\\
b^I_{\m\n}(x,y)&=\sum{\e_{\m\n}}^{\r}X^{II_1}_{\r}(x)Y^{I_1}(y),
&b^I_{ab}(x,y)&=\sum\e_{abc}U^{II_1}(x)\N^c Y^{I_1}(y),\\
b^I_{\m a}(x,y)&=\sum Z_{\m}^{II_3\pm}(x)Y^{I_3\pm}_a(y),
&\phi^{ir}(x,y)&=\sum\phi^{irI_1}(x)Y^{I_1}(y),
\end{align*}
where we have represented
\begin{equation}
h_{ab}=h_{(ab)}+\frac{1}{3}\bar{g}_{ab}h^c_c ,\quad
\bar{g}^{ab}h_{(ab)}=0.
\end{equation}

The various spherical harmonics transform in the following irreducible
representations of $SO(4)\simeq SU(2)_L\times SU(2)_R$:
\begin{itemize}
\item[] Scalar spherical harmonics $Y^I$: $(\frac{k}{2},\frac{k}{2})$, $k\ge0$,
\item[] Vector spherical harmonics $Y_a^{I}=Y_a^{I+}+Y_a^{I-}$:
$(\frac{1}{2}(k+1),\frac{1}{2}(k-1))\oplus(\frac{1}{2}(k-1),
\frac{1}{2}(k+1))$, $k\ge1$,
\item[] Tensor spherical harmonics $Y^I_{(ab)}=Y_{(ab)}^{I+}+Y_{(ab)}^{I-}$:
$(\frac{1}{2}(k+2),\frac{1}{2}(k-2))\oplus(\frac{1}{2}(k-2),
\frac{1}{2}(k+2))$, $k\ge2$.
\end{itemize}
The upper index enumerates a basis in a given
irreducible representation of $SO(4)$:
$I_1=1,\dots,(k+1)^2,\ k\ge0$; $I_3=1,\dots,(k+1)^2-1,\ k\ge1$;
$I_5=1,\dots,(k+1)^2-4,\ k\ge2$.
The action of the Laplacian is
\cite{sal}:
\begin{equation}
\begin{split}
\N^2Y^{I_1} &=-\D Y^{I_1},\\
\N^2Y^{I_3\pm}_a &=(1-\D)Y^{I_3\pm}_a,\qquad\N^aY^{I_3\pm}_a=0,\\
\N^2Y^{I_5\pm}_{(ab)} &=(2-\D)Y^{I_5\pm}_{(ab)},\qquad\N^aY^{I_5\pm}_{(ab)}=0,
\quad\bar{g}^{ab}Y^{I_5\pm}_{(ab)}=0,
\end{split}
\end{equation}
where $\D\equiv k(k+2)$.
The vector spherical harmonics $Y^{I_3\pm}_a$ are also eigenfunctions
of the operator $(*\N)^c_a\equiv{\e_a}^{bc}\N_b$:
\begin{equation}
(*\N)_a^cY^{I_3\pm}_c=\pm(k+1)Y^{I_3\pm}_a.
\end{equation}

We also need to make a number of field redefinitions,
the simplest ones, required to
diagonalize the linearized equations of motion, are
\begin{align}
\phi^{5r}_I &=2ks^r_I+2(k+2)t^r_I,\quad\ U^{r}_I=s^r_I-t^r_I,\\
\pi_I &=-6k\s_I+6(k+2)\t_I,\quad U_I^5=\s_I+\t_I,\\
\label{hmn}h_{\m\n I}&=\varphi_{\m\n I}+\N_{\m}\N_{\n}\zeta_I+g_{\m\n}\eta_I,\\
\label{zeta}\zeta_I&=\frac{4}{k+1}\bigl(\t_I-\s_I\bigr),\\
\label{eta}\eta_I&=\frac{2}{k+1}\bigl(k(k-1)\s_I-(k+2)(k+3)\t_I\bigr),\\
h_{\m I}^{\pm} &=\frac{1}{2}(C_{\m I}^{\pm}-A_{\m I}^{\pm}),
\label{vf}\quad Z_{\m I}^{5\pm}=\pm\frac{1}{4}(C_{\m I}^{\pm}+A_{\m I}^{\pm}).
\end{align}

Here $s_I^r$ and $\sigma_I$ are scalar chiral primaries \cite{deg}.
Note also that we use an {\em off--shell} shift
for $h_{\mu\nu}$ that first appeared in
\cite{gleb7}. It differs from the on--shell shift used in \cite{deg}
by higher order terms.
We recall the origin of the off--shell shift in section 3.3.

The field redefinitions that are needed to make the
equations of motion Lagrangian and to
remove terms with two and four derivatives from the
quadratic corrections to the equations
of motion will be discussed in section 4.
\subsection{Cubic couplings of chiral primaries}
To compute four--point functions involving only chiral primary operators
in the boundary conformal field theory one needs
the quartic couplings giving rise to contact diagrams and cubic couplings
involving at least two chiral primaries,
which contribute to the AdS exchange diagrams.
Here we confine ourselves to the problem of determining the
corresponding cubic couplings.

Obviously, fields like
$\phi^{\underline{i}r}$, $\underline{i}=1,\dots ,4$
that transform as vectors under the $SO(4)$ R--Symmetry
cannot contribute to these couplings.
Therefore, we can set all these fields to zero and, to simplify the notation,
we denote e.g.\ $\phi^{5r}$ as $\phi^r$, etc.

Then the action for the chiral primaries $s^r$ and $\sigma$
may be written in the form
\vspace*{5mm}
\begin{equation}\label{act}
\begin{split}
S(s^r,\s)=\frac{N}{(2\pi)^3}\int d^3x\sqrt{-g_{AdS_3}}&\biggl(
{\mc L}_2(s^r)+{\mc L}_2(t^r)+{\mc L}_2(\s)+{\mc L}_2(\t)+{\mc L}_2(\v^{\pm})\\
&+{\mc L}_2(Z^{r\pm}_{\m})+{\mc L}_2(A_{\m}^{\pm},C_{\m}^{\pm})
+{\mc L}_2(\varphi_{\m\n})\\
&+{\mc L}_3^s(\s)+{\mc L}_3^s(\t)+{\mc L}_3^s(\v^{\pm})+{\mc L}_3^{\s}(\s)
+{\mc L}_3^{\s}(\t)+{\mc L}_3^{\s}(\v^{\pm})\\
&+{\mc L}_3^{s\s}(t^r)+{\mc L}_3^s(A_{\m}^{\pm},C_{\m}^{\pm})
+{\mc L}_3^{\s}(A_{\m}^{\pm},C_{\m}^{\pm})+{\mc L}_3^{s\s}(Z_{\m}^{r\pm})\\
&+{\mc L}_3^s(\varphi_{\m\n})+{\mc L}_3^{\s}(\varphi_{\m\n})\biggr).
\end{split}
\end{equation}
The quadratic terms for the various scalar fields are
\begin{align}
{\mc L}_2(s^r) &=\sum
16k(k+1)\left(-\frac{1}{2}\N_{\m}s^r_I\N^{\m}s^r_I
-\frac{1}{2}m_s^2({s_I^r})^2\right),\\
{\mc L}_2(\s) &=\sum
16k(k-1)\left(-\frac{1}{2}\N_{\m}\s_I\N^{\mu}\s_I
-\frac{1}{2}m_{\s}^2({\s_I})^2\right),\\
{\mc L}_2(t^r) &=\sum
16(k+1)(k+2)\left(-\frac{1}{2}\N_{\m}t^r_I\N^{\m}t^r_I
-\frac{1}{2}m_t^2({t_I^r})^2\right),\\
{\mc L}_2(\t) &=\sum
16(k+2)(k+3)\left(-\frac{1}{2}\N_{\m}\t_I\N^{\m}\t_I
-\frac{1}{2}m_{\t}^2({\t_I})^2\right),\\
{\mc L}_2(\v^{\pm}) &=\sum
\left(-\frac{1}{4}\N_{\m}\v_I^{\pm}\N^{\m}\v_I^{\pm}
-\frac{1}{4}m_{\v}^2({\v_I^{\pm}})^2\right)
\end{align}
with masses
\begin{equation}
m_s^2=m_{\s}^2=k(k-2),\quad m_t^2=m_{\t}^2=(k+2)(k+4),\quad m_{\v}^2=\D.
\end{equation}
The quadratic Lagrangians for the vector fields can be written as
\begin{align}
{\mc L}_2(Z^{r\pm}_{\m}) &=\sum
16(k+1)\left(\mp\frac{1}{4}\e^{\m\n\r}Z^{r\pm}_{\m I}
\partial_{\n}Z^{r\pm}_{\r I}
+\frac{1}{4}m_ZZ^{r\pm}_{\m I}Z^{r\pm\m}_{I}\right) \\
\intertext{for the fields $Z_{\m}^{r\pm}$ with mass $m_Z=k+1$ and}
\label{vector}
{\mc L}_2(A_{\m}^{\pm},C_{\m}^{\pm}) &={\mc L}_2(A_{\m}^{\pm})
+{\mc L}_2(C_{\m}^{\pm})+{\mc L}_2^{\text{cross}}(A_{\m}^{\pm},C_{\m}^{\pm})\\
\intertext{for the fields $A_{\m}^{\pm}$ and $C_{\m}^{\pm}$, where}
{\mc L}_2(A_{\m}^{\pm})&=\sum
\left(-\frac{1}{8}F_{\m\n I}(A^{\pm})F^{\m\n}_I(A^{\pm})
-\frac{1}{4}(k+1)(k-1)A_{\m I}^{\pm}A^{\pm\m}_{I}\mp\frac{1}{2}\e^{\m\n\r}
A_{\m I}^{\pm}\partial_{\n}A_{\r I}^{\pm}\right.\nonumber\\
&\left.\mp\frac{k}{2}P^{\pm}_{k-1}(A^{\pm})^{\m}_IA^{\pm}_{\m I}\right),\\
{\mc L}_2(C_{\m}^{\pm})&=\sum
\left(-\frac{1}{8}F_{\m\n I}(C^{\pm})F^{\m\n}_I(C^{\pm})
-\frac{1}{4}(k+1)(k+3)C_{\m I}^{\pm}C^{\pm\m}_{I}\pm\frac{1}{2}\e^{\m\n\r}
C_{\m I}^{\pm}\partial_{\n}C_{\r I}^{\pm}\right.\nonumber\\
&\left.\mp\frac{1}{2}(k+2)P^{\pm}_{k+3}(C^{\pm})^{\m}_IC^{\pm}_{\m I}\right),\\
{\mc L}_2^{\text{cross}}(A_{\m}^{\pm},C_{\m}^{\pm})&
=\sum\left(\frac{1}{4}F_{\m\n I}(A^{\pm})F^{\m\n}_I(C^{\pm})
-\frac{1}{2}(k-1)(k+3)A_{\m I}^{\pm}C^{\pm\m}_{I}\right.\nonumber\\
&\left.\mp\frac{1}{2}(k+1)\e^{\m\n\r}\bigl(A^{\pm}_{\m I}
\partial_{\n}C^{\pm}_{\r I}
+C^{\pm}_{\m I}\partial_{\n}A^{\pm}_{\r I}\bigr)\right).
\end{align}

Here $F_{\m\n I}(V)=\partial_{\m}V_{\n I}-\partial_{\n}V_{\m I}$ and we
have introduced the first order operators
\begin{equation}
\bigl(P^{\pm}_m\bigr)^{\l}_{\m}\equiv{\e_{\m}}^{\n\l}\N_{\n}\pm
m\d^{\l}_{\m}=\bigl(*\N\bigr)^{\l}_{\m}\pm m\d^{\l}_{\m}.
\end{equation}

Some comments are in order. Since the quadratic action for the vector
fields $Z_{\m}^{r\pm}$ is of the Chern--Simons form it
vanishes on--shell and, therefore,
to compute 2--point functions in the boundary CFT
we have to add certain boundary terms
\cite{gleb5}.
It is also worthwhile to note that the equations
of motion for the vector fields
$A_{\m}^{\pm}$, $C_{\m}^{\pm}$
are not the Proca equations, rather they are Proca--Chern--Simons equations
containing both the usual and the topological mass terms.
Indeed, the equations for  $A_{\m}^{\pm}$ and $C_{\m}^{\pm}$
following directly from (\ref{vector}) are nondiagonal
and both are of the second order.
Adding them produces an equation of the first order
(a constraint) that relates
the fields $A_{\m}^{\pm}$ and $C_{\m}^{\pm}$:
\begin{equation}
\label{cons}
P^{\pm}_{k-1}(A^{\pm})_{\m I}+P^{\pm}_{k+3}(C^{\pm})_{\m I}=0.
\end{equation}
This constraint is then used to obtain the closed Proca--Chern--Simons
equations for the vector fields, e.g.
\begin{equation}
\label{sord}
\N^{\n}F_{\n\m I}(A^{\pm})-(k-1)(k+1)A^{\pm}_{\m I}\mp2{\e_{\m}}^{\n\r}
\partial_{\n}A^{\pm}_{\r I}=P^{\mp}_{k+1}P^{\pm}_{k-1}(A^{\pm})_{\m I}=0.
\end{equation}
Thus, intrinsically one has a second order equation
for one of the vector fields and a constraint
on the second one. The number of physical degrees of freedom described
by a massive pair
$A_{\m I}$, $C_{\m I}$ is then three and this is in agreement with the
discussion in \cite{deg}.
The original equations being components of the second order Einstein equation
and the first order self--duality equation are related to (\ref{cons})
and (\ref{sord})
by simple linear transformations of the fields $A_{\m I}^{\pm}$ and
$C_{\m I}^{\pm}$ ({\it c.f.} \eqref{vf}). Note that the conformal dimensions of the operators in the boundary CFT dual to $A_{\m I}^{\pm}$ and $C_{\m I}^{\pm}$ are $k$, $k+2$ and $k+4$.

Finally, the quadratic Lagrangian for the symmetric second rank tensor field
($\varphi\equiv\varphi^\mu_\mu$) is:
\begin{align}
{\mc L}_2(\varphi_{\m\n}) &=\sum
\left(-\frac{1}{4}\N_{\m}\varphi_{\n\r I}\N^{\m}\varphi^{\n\r}_I
+\frac{1}{2}\N_{\m}
\varphi^{\m\n}_I\N^{\r}\varphi_{\r\n I}
-\frac{1}{2}\N_{\mu}\varphi_I\N_{\n}\varphi^{\n\m}_I
+\frac{1}{4}\N_{\m}\varphi_I\N^{\m}\varphi_I\right.\nonumber\\
&\left.\qquad\ \ +\frac{1}{4}(2-\D)\bigl(\varphi_{\m\n I}\bigr)^2
+\frac{1}{4}\D\bigl(\varphi_I\bigr)^2\right).
\end{align}

The cubic couplings of scalar fields are
\begin{equation}
{\mc L}_3^s(\psi)=V^{ss\psi}_{I_1I_2I_3}s_{I_1}^rs_{I_2}^r\psi_{I_3},\quad
{\mc L}_3^{\s}(\psi)=V^{\s\s\psi}_{I_1I_2I_3}\s_{I_1}\s_{I_2}\psi_{I_3},\quad
{\mc L}_3^{s\s}(t^r)=V^{st\s}_{I_1I_2I_3}s_{I_1}^rt_{I_2}^r\s_{I_3},
\end{equation}
with $\psi\in\{\s,\t,\v^{\pm}\}$ and the vertices
(the notation is explained in eqs.(\ref{notation1}) and (\ref{notation2}))
\begin{align}
\label{vertex1}
V^{ss\s}_{I_1I_2I_3}&=
-\frac{2^4(\S-2)\S(\S+2)\a_1\a_2\a_3}{k_3+1}a_{I_1I_2I_3},\\
V^{ss\t}_{I_1I_2I_3}&=
\frac{2^6(\S+2)(\a_1+1)(\a_2+1)\a_3(\a_3-1)(\a_3-2)}{k_3+1}a_{I_1I_2I_3},\\
\label{coupl}
V^{\s\s\s}_{I_1I_2I_3}&=
-\frac{2^3(\S-2)\S(\S+2)\a_1\a_2\a_3}{3(k_1+1)(k_2+1)(k_3+1)}
\bigl(k_1^2+k_2^2+k_3^2-2\bigr)a_{I_1I_2I_3},\\
V^{\s\s\t}_{I_1I_2I_3}&=
\frac{2^5(\S+2)(\a_1+1)(\a_2+1)\a_3(\a_3-1)(\a_3-2)}{(k_1+1)(k_2+1)(k_3+1)}
\bigl(k_1^2+k_2^2+(k_3+2)^2-2\bigr)a_{I_1I_2I_3},
\end{align}
\begin{align}
V^{st\s}_{I_1I_2I_3}&=
\frac{2^7(\S+2)(\a_1+1)\a_2(\a_2-1)(\a_2-2)(\a_3+1)}{k_3+1}a_{I_1I_2I_3},\\
V^{ss\v^{\pm}}_{I_1I_2I_3}&=
2^2\S(\a_3-1)p_{I_1I_2I_3}^{\pm},\\
V^{\s\s\v^{\pm}}_{I_1I_2I_3}&=
\frac{2\S(\a_3-1)}{(k_1+1)(k_2+1)}\bigl(k_1^2+k_2^2-(k_3+1)^2-1
\bigr)p_{I_1I_2I_3}^{\pm}.
\end{align}
In our notation, the vertices \eqref{vertex1} and \eqref{coupl} are precisely the ones
found in \cite{mih}.

Cubic terms involving two chiral primaries and the vector fields
$A_{\m}^{\pm}$, $C_{\m}^{\pm}$ can be represented as
\begin{align}
\label{AC1}
{\mc L}_3^s(A_{\m}^{\pm},C_{\m}^{\pm})&=
V^{ssA^{\pm}}_{I_1I_2I_3}s_{I_1}^r\N^{\m}s_{I_2}^rA_{\m
I_3}^{\pm}+
V^{ssC^{\pm}}_{I_1I_2I_3}s_{I_1}^r\N^{\m}s_{I_2}^rC_{\m I_3}^{\pm}\\
&\pm W^{s\pm}_{I_1I_2I_3}
\bigl(P^{\pm}_{k_3-1}(s_{I_1}^r\N s_{I_2}^r)^{\m}A_{\m I_3}^{\pm}
-P^{\pm}_{k_3+3}(s_{I_1}^r\N s_{I_2}^r)^{\m}C_{\m I_3}^{\pm}\bigr),\nonumber\\
\label{AC2}
{\mc L}_3^{\s}(A_{\m}^{\pm},C_{\m}^{\pm})&= V^{\s\s
A^{\pm}}_{I_1I_2I_3}\s_{I_1}\N^{\m}\s_{I_2}A_{\m I_3}^{\pm}+
V^{\s\s C^{\pm}}_{I_1I_2I_3}\s_{I_1}\N^{\m}\s_{I_2}C_{\m I_3}^{\pm}\\
&\pm W^{\s\pm}_{I_1I_2I_3}
\bigl(P^{\pm}_{k_3-1}(\s_{I_1}\N\s_{I_2})^{\m}A_{\m I_3}^{\pm}
-P^{\pm}_{k_3+3}(\s_{I_1}\N\s_{I_2})^{\m}C_{\m I_3}^{\pm}\bigr),\nonumber\\
\intertext{whereas the interaction of $s^r$ and $\s$
with the fields $Z_{\m}^{r\pm}$ is found to be}
{\mc L}_3^{\s s}(Z_{\m}^{r\pm})&=
\pm V^{\s sZ^{\pm}}_{I_1I_2I_3}\s_{I_1}\N^{\mu}s^r_{I_2}Z^{r\pm}_{\m I_3}.
\end{align}
These expressions describe the minimal interactions of
vector fields with two scalars in three dimensions.
Here the couplings are
\begin{align}
V^{ssA^{\pm}}_{I_1I_2I_3}&=-2(\S+1)(\S-1)t_{I_1I_2I_3}^{\pm},\\
V^{ssC^{\pm}}_{I_1I_2I_3}&=2(2\a_3-1)(2\a_3-3)t_{I_1I_2I_3}^{\pm},\\
W^{s\pm}_{I_1I_2I_3}&=2(k_3+1)t_{I_1I_2I_3}^{\pm},\\
V_{I_1I_2I_3}^{\s\s A^{\pm}}&=-\frac{(\S+1)(\S-1)}{(k_1+1)(k_2+1)}
\bigl(k_1^2+k_2^2-k_3^2-1\bigr)t_{I_1I_2I_3}^{\pm},\\
V_{I_1I_2I_3}^{\s\s C^{\pm}}&=\frac{(2\a_3-1)(2\a_3-3)}{(k_1+1)(k_2+1)}
\bigl(k_1^2+k_2^2-(k_3+2)^2-1\bigr)t_{I_1I_2I_3}^{\pm},\\
W^{\s\pm}_{I_1I_2I_3}&=-2(k_3+1)\frac{(k_1-1)(k_2-1)}{(k_1+1)(k_2+1)}
t_{I_1I_2I_3}^{\pm},\\
V_{I_1I_2I_3}^{\s sZ^{\pm}}&=\frac{2^4(\S+1)(2\a_3-1)(k_3+1)}{k_1+1}
t_{I_1I_2I_3}^{\pm}.
\end{align}

Finally the interaction of chiral primaries with symmetric
tensors of the 2nd rank are
\begin{align}
{\mc L}_3^s(\varphi_{\m\n})&=V^{ss\varphi}_{I_1I_2I_3}
\left(\N^{\m}s_{I_1}^r\N^{\n}s_{I_2}^r\varphi_{\m\n I_3}
-\frac{1}{2}\bigl(\N^{\m}s_{I_1}^r\N_{\m}s_{I_2}^r
+\frac{1}{2}(m_1^2+m_2^2-\D_3)s_{I_1}^rs_{I_2}^r\bigr)\varphi_{I_3}\right),\\
{\mc L}_3^{\s}(\varphi_{\m\n})&=V^{\s\s\varphi}_{I_1I_2I_3}
\left(\N^{\m}\s_{I_1}\N^{\nu}\s_{I_2}\varphi_{\m\n I_3}
-\frac{1}{2}\bigl(\N^{\m}\s_{I_1}\N_{\m}\s_{I_2}
+\frac{1}{2}(m_1^2+m_2^2-\D_3)\s_{I_1}\s_{I_2}\bigr)\varphi_{I_3}\right),
\end{align}
where
\begin{align}
V^{ss\varphi}_{I_1I_2I_3}&=2^2(\S+2)\a_3a_{I_1I_2I_3},\\
V^{\s\s\varphi}_{I_1I_2I_3}&=\frac{2(\S+2)\a_3}{(k_1+1)(k_2+1)}
\bigl(k_1^2+k_2^2-(k_3+1)^2-1\bigr)a_{I_1I_2I_3}.
\end{align}
Above the summation over $I_1$, $I_2$, $I_3$ and $r$ is
assumed and we have defined
\begin{equation}
\label{notation1}
\S\equiv k_1+k_2+k_3,\quad\a_i\equiv\frac{1}{2}(k_l+k_m-k_i),\
l\neq m\neq i\neq l,
\end{equation}
and ({\it c.f.} also Appendix B)
\begin{equation}
\label{notation2}
a_{I_1I_2I_3}\equiv\int Y_{I_1}Y_{I_2}Y_{I_3},\quad
t_{I_1I_2I_3}^{\pm}\equiv\int\N^aY_{I_1}Y_{I_2}Y_{aI_3}^{\pm},\quad
p_{I_1I_2I_3}^{\pm}\equiv\int\N^aY_{I_1}\N^bY_{I_2}Y_{(ab)I_3}^{\pm}.
\end{equation}
To summarize, we have the following types of cubic vertices:\\
\begin{picture}(50000,23000)
\gaplength=250
\drawline\scalar[\SE\REG](2000,20000)[3]
\put(3000,\pfronty){$s^r$, $\s$}
\gaplength=250
\drawline\scalar[\E\REG](\pbackx,\pbacky)[3]
\put(\pbackx,\pbacky){\ $\s$, $\t$, $\v^{\pm}$}
\gaplength=250
\drawline\scalar[\SW\REG](\pfrontx,\pfronty)[3]
\put(3000,\pbacky){$s^r$, $\s$}
\gaplength=250
\drawline\scalar[\SE\REG](18000,20000)[3]
\put(19000,\pfronty){$s^r$}
\gaplength=250
\drawline\scalar[\E\REG](\pbackx,\pbacky)[3]
\put(\pbackx,\pbacky){\ $t^r$}
\gaplength=250
\drawline\scalar[\SW\REG](\pfrontx,\pfronty)[3]
\put(19000,\pbacky){$\s$}
\gaplength=250
\drawline\scalar[\SE\REG](2000,10000)[3]
\put(3000,\pfronty){$s^r$, $\s$}
\drawline\photon[\E\REG](\pbackx,\pbacky)[5]
\put(\pbackx,\pbacky){\ $A_{\m}^{\pm}$, $C_{\m}^{\pm}$}
\gaplength=250
\drawline\scalar[\SW\REG](\pfrontx,\pfronty)[3]
\put(3000,\pbacky){$s^r$, $\s$}
\gaplength=250
\drawline\scalar[\SE\REG](18000,10000)[3]
\put(19000,\pfronty){$s^r$}
\drawline\photon[\E\REG](\pbackx,\pbacky)[5]
\put(\pbackx,\pbacky){\ $Z^{r\pm}_{\m}$}
\gaplength=250
\drawline\scalar[\SW\REG](\pfrontx,\pfronty)[3]
\put(19000,\pbacky){$\s$}
\gaplength=250
\drawline\scalar[\SE\REG](32000,15000)[3]
\put(33000,\pfronty){$s^r$, $\s$}
\drawline\gluon[\E\CENTRAL](\pbackx,\pbacky)[5]
\put(\pbackx,\pbacky){\ $\varphi_{\m\n}$}
\gaplength=250
\drawline\scalar[\SW\REG](\pfrontx,\pfronty)[3]
\put(33000,\pbacky){$s^r$, $\s$}
\put(2000,0){Table 1: Cubic vertices containing two supergravity
fields dual to CPOs.}
\end{picture}\vspace*{2mm}
In particular we see that all possible cubic invariants under
$SO(n)\times SO_R(4)$ containing at least two supergravity
fields dual to CPOs are present.

\subsection{Cubic couplings at extremality} With the cubic
couplings at hand the problem of computing the 3--point
correlation functions of two CPOs with an operator associated to
another gravity field entering the cubic vertex becomes
straightforward. One needs to determine the on--shell value of
the corresponding cubic action, which amounts to computing
certain integrals over the AdS space, where for the latter
problem a well--developed technique is available \cite{fr}.
Generally the AdS integrals diverge for some ``extremal'' values
of conformal dimensions (masses) of the fields involved and this
is an indication that the corresponding supergravity coupling
should vanish, otherwise the correlation function would be
ill--defined \cite{fr}, \cite{sei}. For example, the AdS integral
corresponding to the 3--point correlation function of scalar
fields with conformal dimensions $\D_1$,  $\D_2$ and $\D_3$ is
ill--defined if $\D_1+\D_2=\D_3$ (or any relation obtained from
this by permutation of indices). Inspection shows that the cubic
couplings we found do indeed vanish at extremality, i.e.\ when
the accompanying AdS integrals diverge. The only case where this
property can not be seen straightforwardly is for the couplings
of scalar fields with vector fields $A_{\m}^{\pm}$ and
$C_{\m}^{\pm}$. Below we present the analysis making the property
of vanishing at extremality manifest.

Recall that due to \eqref{cons} the fields $A_{\m}^{\pm}$ and
$C_{\m}^{\pm}$ do not describe independent degrees of freedom.
Regarding, {\it e.g.}, $A_{\m}^{\pm}$ as independent variables
we first consider the solution of eq.\eqref{sord} satisfying
$$
P^{\pm}_{k-1}(A^{\pm})_{\m I}=0.
$$
Then the constraint \eqref{cons} gives $P^{\pm}_{k+3}(C^{\pm})_{\m
I}=0$. Clearly, the last two equations imply the Maxwell equations
\begin{eqnarray}
\N^{\n}F_{\n\m I}(A^{\pm})-(k-1)^2A^{\pm}_{\m I}=0,~~~~
\N^{\n}F_{\n\m I}(C^{\pm})-(k+3)^2C^{\pm}_{\m I}=0.
\end{eqnarray}
Therefore, the masses of the vector fields $A^{\pm}_{\m I}$ and
$C^{\pm}_{\m I}$ are $m_A=k-1$ and $m_C=k+3$. Recalling the
formula for the conformal weight $\D_V$ of an operator dual to a
vector field $V_{\m}$ with mass $m$ in $AdS_{d+1}$ (see, {\it e.g.}
\cite{aha}) we find $\D_A=k$ and $\D_C=k+4$. It is worthwhile to note that
$\D_A$ has the same conformal dimension as the scalar CPOs. The corresponding CFT operators are the vector CPOs in the spin 2 tower of supermultiplets \cite{deg}. 

The evaluation of the 3--point functions of CPOs with
vector fields requires the knowledge of the following AdS integral
\begin{equation}
\nonumber
\int\frac{d^3\o}{\o_0^3}K_{\D_1}(\o,\x_1)\N^{\m}K_{\D_2}(\o,\x_2){\mc
G}_{\m i\,\D_3}(\o,\x_3)=
\frac{R_{123}}{|\x_{12}|^{\D_1+\D_2-\D_V}
|\x_{13}|^{\D_1+\D_V-\D_2}|\x_{23}|^{\D_2+\D_V-\D_1}}\frac{Z_i}{Z},
\end{equation}
where the coordinate dependence on the r.h.s. is completely fixed
by the conformal symmetry. Here $\x_i$ are the positions of the
operators in the correlation function of the boundary CFT,
$\x_{ij}=\x_i-\x_j$,
\begin{equation*}
Z_i=\frac{(\x_{13})_i}{\x_{13}^2}-\frac{(\x_{23})_i}{\x_{23}^2},~~~~
Z^2=Z_iZ_i
\end{equation*}
and $K_{\D}(\o,\x)$, ${\mc G}_{\m\n\,\D_3}(\o,\x)$ are the scalar
and vector bulk--to--boundary propagators respectively. Applying
the inversion method \cite{fr} one finds for $R_{123}$ the
following answer:
\begin{align}
R_{123}=&\frac{1}{\pi^2}\frac{\Gamma\bigl(\frac{1}{2}(\D_1+\D_2-\D_V+1)\bigr)\Gamma\bigl(\frac{1}{2}(\D_1+\D_V-\D_2+1)\bigr)\Gamma\bigl(\frac{1}{2}(\D_2+\D_V-\D_1+1)\bigr)}{\Gamma(\D_1-1)\Gamma(\D_2-1)\Gamma(\D_V)}\nonumber\\
&\times\Gamma\bigl(\textstyle{\frac{1}{2}}(\D_1+\D_2+\D_V-1)\bigr).
\end{align}
$R_{123}$ is ill--defined in several cases. First we consider the case when\footnote{$R_{123}$ is also divergent for $\D_1+\D_2-\D_V+1$ a negative integer, but in that cases $t^{\pm}_{I_1I_2I_3}=0$.}
\begin{equation}
\label{D1}
\D_1+\D_2-\D_V+1=0.
\end{equation}
For CPOs with $\D=k$ this equation becomes $k_1+k_2-\D_V+1=0$ and,
therefore, for $C^{\pm}_{\m}$ it reads as
$$
k_1+k_2-\D_C+1=k_1+k_2-k_3-3=0,
$$
i.e., $\a_3=3/2$. But the couplings $V^{ss C^{\pm}}_{I_1I_2I_3}$ and $V^{\s\s
C^{\pm}}_{I_1I_2I_3}$ (see \eqref{AC1} and \eqref{AC2}) contain the factor
$2\a_3-3$ and, therefore, vanish.\footnote{The terms in
\eqref{AC1} and \eqref{AC2} proportional to $W^{s\pm}_{I_1I_2I_3}$ and to
$W^{\s\pm}_{I_1I_2I_3}$ vanish after integrating by parts and taking into
account the equations of motion for $A_{\m}^{\pm}$ and
$C_{\m}^{\pm}$.}  Ccomputing the correlation functions involving
the fields $A^{\pm}_{\m}$ a divergence arises when
$$
k_1+k_2-\D_A+1=k_1+k_2-k_3+1=0,
$$
i.e., when $\a_3=-1/2$. However, the couplings  $V^{ssA^{\pm}}_{I_1I_2I_3}$
and $V^{\s\s A^{\pm}}_{I_1I_2I_3}$ contain the tensors $t_{I_1I_2I_3}^{\pm}$ that are non--vanishing only if $k_1+k_2\geq k_3+1$ (and relations
obtained by permutation of the indices). Hence, the divergence is
irrelevant since the couplings are zero due to the vanishing of
$t_{I_1I_2I_3}^{\pm}$.

Moreover, $R_{123}$ also diverges when
\begin{equation}
\label{D2} \D_1+\D_V-\D_2+1=0.
\end{equation}
For $C^{\pm}_{\m}$ this gives $k_1+k_3=k_2-5$, i.e., $\a_2=-5/2$.
On the other hand, non--vanishing of $t_{I_1I_2I_3}^{\pm}$ requires the
inequality $k_1+k_3\geq k_2+1$, so that for the case under
consideration $t_{I_1I_2I_3}^{\pm}$ again vanish. For $A^{\pm}_{\m}$
eq.(\ref{D2}) gives $k_1+k_3=k_2-1$, i.e., $\a_2=-1/2$, and the
couplings vanish by the same reason as for $C_{\m}^{\pm}$.

Equation \eqref{sord} has another solution obeying
$P^{\mp}_{k+1}(A^{\pm})_{\m I}=0$, which we now consider. Perform
the shift
\begin{equation}
\label{shift} C^{\pm}_{\m I}= {C'}^{\pm}_{\m
I}-\frac{k}{k+2}A^{\pm}_{\m I},
\end{equation}
where $A^{\pm}_{\m I}$ is not arbitrary, rather it solves
$P^{\mp}_{k+1}(A^{\pm})_{\m I}=0$. Then the linear constraint
\eqref{cons} turns into
$$
P^{\pm}_{k+3}({C'}^{\pm})_{\m
I}+\frac{2}{k+2}P^{\mp}_{k+1}(A^{\pm})_{\m
I}=P^{\pm}_{k+3}({C'}^{\pm})_{\m I}=0.
$$
Thus, ${C'}^{\pm}_{\m}$ decouple from $A^{\pm}_{\m}$. The fields
$A^{\pm}_{\m}$ then correspond to operators with $\D_A=k+2$. The
divergence \eqref{D1} now gives $k_1+k_2=k_3+1$, i.e.,
$\a_3=1/2$. The coupling of two scalars with the vector fields
$A^{\pm}_{\m}$ corrected by the shift \eqref{shift} (we again
integrate the terms in \eqref{AC1} and \eqref{AC2} proportional
to $W^{s\pm}_{I_1I_2I_3}$ and to $W^{\s\pm}_{I_1I_2I_3}$ 
by parts and use the equations of
motion for $A_{\m}^{\pm}$ and ${C'}_{\m}^{\pm}$) reads 
\begin{equation}
\bar{V}^{\s\s A^{\pm}}_{I_1I_2I_3}\equiv V^{\s\s
A^{\pm}}_{I_1I_2I_3}+V^{\s\s
C^{\pm}}_{I_1I_2I_3}+4k_3W^{\s\pm}_{I_1I_2I_3}
\end{equation}
and analogously for $s^r$. The explicit results are given by
\begin{align}
\bar{V}^{ss A^{\pm}}_{I_1I_2I_3}&=-8(k_3+1)(2\a_3-1)t^{\pm}_{I_1I_2I_3},\\
\bar{V}^{\s\s
A^{\pm}}_{I_1I_2I_3}&=-4(k_3+1)(2\a_3-1)\frac{(k_1+1)(k_1+k_3)+(k_2+1)(k_2+k_3)-4(k_3+1)}{(k_1+1)(k_2+1)}t^{\pm}_{I_1I_2I_3}
\end{align}
and vanish at extremality. The AdS integral is also divergent for
\eqref{D2}, i.e. for  $\a_1=-3/2$. However, in
this case $t^{\pm}_{I_1I_2I_3}$ is zero.

Thus, we have shown that all the cubic couplings we determined vanish in
the extremal cases.

\subsection{Truncation to the graviton multiplet}
The bosonic part of the Lagrangian density for the
three--dimensional supergravity based on
the $SU(1,1|2)_L\times SU(1,1|2)_R$ supergroup is \cite{at}
\begin{equation}
\label{adsSUGRA}
{\mc L}=R+2-\e^{\m\n\r}\bigl(A_{\m}^{ij}\partial_{\n}A_{\r}^{ji}
+\frac{2}{3}A_{\m}^{ij}A_{\n}^{jk}A_{\r}^{ki}\bigr)
+\e^{\m\n\r}\bigl({A'}_{\m}^{ij}\partial_{\n}{A'}_{\r}^{ji}
+\frac{2}{3}{A'}_{\m}^{ij}{A'}_{\n}^{jk}{A'}_{\r}^{ki}\bigr),
\end{equation}
where $A_{\m}^{ij}=-A_{\m}^{ji}$, ${A'}_{\m}^{ij}=-{A'}_{\m}^{ji}$
are the $SO(3)$ gauge fields and according
to our conventions we have set the cosmological constant to $-1$.

We now demonstrate that the lowest modes of the vector fields
$A_{\m}^{\pm}$ obey the first order Chern--Simons equations,
although generically
the equations of motion are of second order. Thus,
we consider the self--interaction of the vector fields $A_{\m}^{\pm}$
and restrict ourselves to the case where two of
the three fields, say $A_{\m I_2}^{\pm}$, $A_{\m I_3}^{\pm}$
come from the massless
graviton multiplet, i.e.\ their equations of motion are
\begin{equation}
P_0(A^{\pm})_{\m}={\e_{\m}}^{\n\r}\partial_{\n}A_{\r}^{\pm}=0\
\Longleftrightarrow\ \N_{\m}A_{\n}^{\pm}=\N_{\n}A_{\m}^{\pm}.
\end{equation}
Then the quadratic corrections to the linear constraint
(\ref{cons}) can be written as
\begin{equation}
P^{\pm}_{k_1-1}(A^{\pm})_{\m I_1}+P^{\pm}_{k_1+3}(C^{\pm})_{\m I_1}=
\pm \frac{1}{2}{\e_{\m}}^{\n\r}A^{\pm}_{\n I_2}A^{\pm}_{\r I_3}
\int\e^{abc}Y_{a I_1}^{\pm}Y_{b I_2}^{\pm}Y_{c I_3}^{\pm}.
\end{equation}
Since both vector fields on the r.h.s. transform in the
$(1,0)$ of $SU(2)_L\times SU(2)_R$ (or $(0,1)$ respectively),
$Y_{b I_2}^{\pm}Y_{c I_3}^{\pm}$ transform as
\begin{equation}
(1,0)\otimes(1,0)=(0,0)\oplus(1,0)\oplus(2,0);
\qquad(0,1)\otimes(0,1)=(0,0)\oplus(0,1)\oplus(0,2),
\end{equation}
and therefore the $S^3$ integral is nonzero only if $k_1=1$.
In this case we have
\begin{equation}
P_0(A^{\pm})_{\m I_1}+P^{\pm}_{4}(C^{\pm})_{\m I_1}=
\pm \frac{1}{2}{\e_{\m}}^{\n\r}A^{\pm}_{\n I_2}A^{\pm}_{\r I_3}
\int\e^{abc}Y_{a I_1}^{\pm}Y_{b I_2}^{\pm}Y_{c I_3}^{\pm}.
\end{equation}
On the other hand, it is easy to show that there is no coupling of
$C^{\pm}_{\m I}$ with two massless vector fields and therefore
it is consistent to set the fields $C_{\m I}^{\pm}$ to zero.

Since the $S^3$ integral is completely antisymmetric in
$I_1$, $I_2$ and $I_3$ (and the $I_i$ run from 1 to 3) it is proportional to
$\e^{I_1I_2I_3}$ and can be represented as
$\mp2C_{\pm I_1}^{ij}C_{\pm I_2}^{jk}C_{\pm I_3}^{ki}$,
where $C_{\pm I}^{ij}=-C_{\pm I}^{ji}$.
Defining
\begin{equation}
A_{\m}^{ij}=C^{ij}_{+I}A_{\m}^{I+},\qquad{A'}_{\m}^{ij}=C^{ij}_{-I}A_{\m}^{I-}
\end{equation}
the equation of motion for $A_{\m}^{ij}$ reads
\begin{equation}
{\e_{\m}}^{\n\r}\partial_{\n}A_{\r}^{ij}
=-{\e_{\m}}^{\n\r}A_{\n}^{ik}A_{\r}^{kj}
\end{equation}
and analogously for ${A'}_{\m}^{ij}$.
These are precisely the equations of motion following
from eq.(\ref{adsSUGRA}).

Now we address the issue of the consistency of the KK truncation
to the sum of two multiplets, one of them naturally the massless
graviton multiplet and a second one containing lowest mode scalar
CPOs. Surprisingly, all the cubic couplings we computed involving two fields
from the sum of the massless graviton multiplet and the special
spin--$\frac{1}{2}$ multiplet\footnote{Generically the multiplets in the vector representation of $SO(n)$ involve fields with spin 1. However at the lowest level these are absent \cite{deg}.}
and one field belonging to any
other multiplet vanish.\footnote{For the cubic couplings with
vector fields see section 2.2.} Recall that the
spin--$\frac{1}{2}$ multiplet contains the scalar modes $s^r$ with
$k=1$ and $\phi^{\underline{i}r}$ with $k=0$, and
spin--$\frac{1}{2}$ states $\chi^r$ \cite{deg}. All the operators
in the boundary CFT dual to the gravity fields from the
spin--$\frac{1}{2}$ multiplet are either relevant or marginal.
Based on the analysis presented here, one cannot exclude that
a consistent truncation to the sum ``massless graviton multiplet
$+$ special spin--$\frac{1}{2}$ multiplet'' does exist. Of course, 
only on the basis of the cubic vertices considered here, this issue 
cannot be decided. 
It is worthwhile to  note that $s^r$ with $k=1$ correspond in the
boundary CFT to the scalar CPOs with the lowest conformal
dimension.

Another natural example to consider is 
the lowest level of the spin 1 $SO(n)$ singlet multiplet, containing $\s$ with 
$k=2$.
Here, however, the consistent truncation is not
possible. Indeed, the cubic coupling of two CPOs and one symmetric
second rank (massive) tensor
\begin{equation}
V^{\s\s\varphi}_{I_1I_2I_3}\sim(\S+2)\a_3
\bigl(k_1^2+k_2^2-(k_3+1)^2-1\bigr)a_{I_1I_2I_3}
\end{equation}
does not vanish if $k_1=k_2=k_3=2$. Note also that the CFT
multiplet dual to the $SO(n)$ singlet discussed above contains
irrelevant operators.

In the following two sections we describe how to obtain the cubic
action \eqref{act} from the covariant equations of motion \eqref{eqn}.
\section{The linearized equations of motion}
In this section we discuss the linearized equations
of motion for physical fields.
We also keep track of the corrections ${\mc Q}$ to the
linearized equations though for the sake of
clarity we do not give their explicit expressions. They can be found in \cite{ari}.
The structure of the quadratic
corrections will be discussed in section 4.
\subsection{Scalar fields}
Scalar fields arise from two sources: from
the equations of motion for the scalars $\phi^r$ and the
2--forms $B^r$, where $r=1,\dots ,n$, and
from the sphere components of the Einstein equation and the
equation for the 2--form $B$. We start with the first category and obtain
\begin{align}
\label{1}
&(\N_{\m}^2+\N_a^2)\phi^r+4(\N_{\m}^2+\N_a^2)U^r
=4\N^{\m}(\N_{\m}U^r-X_{\m}^r)+{\mc Q}_1^r,\\
\intertext{from the equation of motion for $\phi^r$,} \label{con1}
&\N_a(X_{\m}^r-\N_{\m}U^r)+{\e_a}^{bc}\N_bb^r_{\m c}
-{\e_{\m}}^{\n\r}\N_{\n}b^r_{\r a}={\mc Q}_{2a\m}^r\\ \intertext{and}
&(\N_{\m}^2+\N_a^2)U^r+2\phi^r=\N^{\m}(\N_{\m}U^r-X_{\m}^r)+{\mc Q}_3^r
\label{2}
\end{align}
from the $(\m\n a)$ and $(abc)$ components of the
antiself--duality equation for $B^r$. ${\mc Q}_i^r$ denote higher
order corrections to the linearized equations.

First we expand all the fields in spherical harmonics on $S^3$.
Equation \eqref{con1} contains both the transversal
terms proportional to $Y_{a}^{I_3\pm}$
and the longitudinal ones proportional to $\N_aY^{I_1}$.
Projecting \eqref{con1} on the longitudinal part by multiplying it with $\N^a$
one gets a linear constraint. One then uses this constraint to diagonalize
the system \eqref{1} and \eqref{2}. The resulting equations are
\begin{equation}
\begin{split}
4(k+1)\left(\N_{\m}^2-k(k-2)\right)s^{rI_1}&={\mc Q}_s^{rI_1},\\
4(k+1)\left(\N_{\m}^2-(k+2)(k+4)\right)t^{rI_1}&={\mc Q}_t^{rI_1},
\end{split}
\end{equation}
where
\begin{equation}
{\mc Q}_s^{rI_1}={\mc Q}_1^{rI_1}+\frac{2}{k}\N^{\m}\N^a{\mc Q}_{2a\m}^{rI_1}
+2k{\mc Q}_3^{rI_1},\quad
{\mc Q}_t^{rI_1}={\mc Q}_1^{rI_1}-\frac{2}{k+2}\N^{\m}\N^a{\mc Q}_{2a\m}^{rI_1}
-2(k+2){\mc Q}_3^{rI_1}
\end{equation}
and
\begin{align}
s^r_{I_1} & = \frac{1}{4(k+1)}\bigl(\phi^r_{I_1}+2(k+2)U^r_{I_1}\bigr), \\
\intertext{for the chiral primary $s^r$ and} t^r_{I_1} & =
\frac{1}{4(k+1)}\bigl(\phi^r_{I_1}-2kU^r_{I_1}\bigr)
\end{align}
for the scalar $t^r$ ({\it c.f.} (2.17)).

For the second category we obtain
\begin{align}
-\frac{1}{6}\left(\N_{\m}^2+\N_a^2-8\right)h_b^b-4\N_a^2U-\frac{1}{6}\N_a^2
\bigl(h^{\m}_{\m}
+\frac{1}{3}h_b^b\bigr)&={\mc Q}_1\\ \intertext{and}
\label{con2}-\frac{1}{2}\left(\N_{\m}^2+\N_a^2-2\right)h_{(ab)}
-\frac{1}{2}\N_{(a}\N_{b)}\bigl(h^{\m}_{\m}+\frac{1}{3}h_c^c\bigr)
+\N_{(a}\N^{\m}h_{b)\m}&={\mc Q}_{2(ab)}
\\ \intertext{from the $(ab)$ components of the
Einstein equation together with}
\label{con3}\N_a\left(X_{\m}+\nabla_{\m}U\right)-{\e_a}^{bc}\N_bb_{\m c}
-{\e_{\m}}^{\n\r}\N_{\n}b_{\r a}-h_{a\m}&={\mc Q}_{3a\m} \\ \intertext{and}
\left(\N_{\m}^2+\N_a^2\right)U-\frac{2}{3}h_b^b+\frac{1}{2}\bigl(h^{\m}_{\m}
+\frac{1}{3}h_b^b\bigr)
-\N^{\m}\left(X_{\m}+\N_{\m}U\right)&={\mc Q}_4
\end{align}
from the $(\m\n a)$ and $(abc)$ components of the self--duality
equation for $B$.

Again, expanding the fields in spherical harmonics, solving
the longitudinal constraints \eqref{con2}
(i.e.\ multiplying \eqref{con2} by $\N^a\N^b$) and \eqref{con3}
and diagonalizing the resulting system of equations we obtain
\begin{equation}
\begin{split}
2(k+1)\left(\N_{\m}^2-k(k-2)\right)\s^{I_1}&={\mc Q}_{\s}^{I_1} ,\\
2(k+1)\left(\N_{\m}^2-(k+2)(k+4)\right)\t^{I_1}&={\mc Q}_{\t}^{I_1},
\end{split}
\end{equation}
where
\begin{align}
{\mc Q}_{\s}^{I_1}&={\mc Q}_1^{I_1}
+\frac{1}{2k(k-1)}\N^a\N^b{\mc Q}_{2(ab)}^{I_1}
-\frac{1}{k}\N^{\m}\N^a{\mc Q}_{3a\m}^{I_1}+(k+2){\mc Q}_4^{I_1}, \\
{\mc Q}_{\t}^{I_1}&=-{\mc Q}_1^{I_1}
-\frac{1}{2(k+2)(k+3)}\N^a\N^b{\mc Q}_{2(ab)}^{I_1}
-\frac{1}{k+2}\N^{\m}\N^a{\mc Q}_{3a\m}^{I_1}+k{\mc Q}_4^{I_1}
\end{align}
and the chiral primary field $\s_{I_1}$ and the scalar
$\t_{I_1}$ are defined as
({\it c.f.} (2.18)):
\begin{align}
\s_{I_1}&=\frac{1}{12(k+1)}\bigl(6(k+2)U_{I_1}-\pi_{I_1}\bigr), \\
\t_{I_1}&=\frac{1}{12(k+1)}\bigl(6kU_{I_1}+\pi_{I_1}\bigr).
\end{align}

Finally, the equations of motion for the scalars $\v^{I_5\pm}$ originate
from the transverse traceless part of eq.\eqref{con2}
\begin{equation}
-\frac{1}{2}\left(\N_{\m}^2-\D\right)\v^{I_5\pm}={\mc Q}_2^{I_5\pm}.
\end{equation}
We summarize the results for the relevant scalar
modes ({\it c.f.} Table 1) in Fig.\ 1:
\vskip 0.2cm
\[ \beginpicture \setquadratic
\setcoordinatesystem units <1cm,1.7mm>
\setplotarea x from 0 to 5, y from -1 to 27
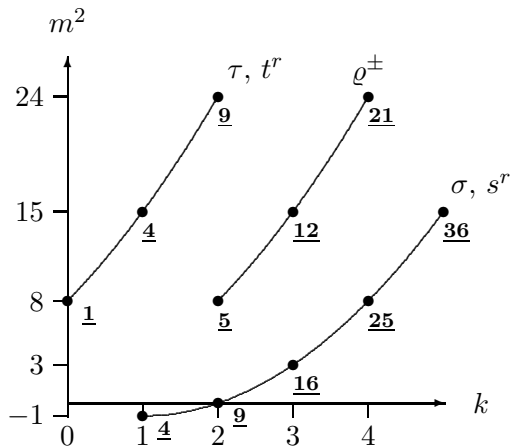
\axis bottom shiftedto y=0 label {Fig.\ 1: Mass spectrum of scalars.}
ticks numbered from 0 to 4 by 1 /
\axis left ticks withvalues $-1$  $3$ $8$ $15$ $24$ {} /
at -1 3 8 15 24 / /
\put {\vector(1,0){60}} [Bl] at 5 0
\put {$k$} at 5.5 0
\put {\vector(0,1){60}} [Bl] at 0 27
\put {$m^2$} at 0 30
\plot 2 8  2.5 11.25  3 15  3.5 19.25  4 24 /
\put {$\v^{\pm}$} at 4 26
\unitlength1mm
\put {\circle*{1.5}} [Bl] at 2 8
\put {\circle*{1.5}} [Bl] at 3 15
\put {\circle*{1.5}} [Bl] at 4 24
\put {\scriptsize{${\bf\underline{5}}$}} [Bl] at 2 6
\put {\scriptsize{${\bf\underline{12}}$}} [Bl] at 3 13
\put {\scriptsize{${\bf\underline{21}}$}} [Bl] at 4 22
\plot 1 -1  2 0  3 3  3.5 5.25  4 8  4.5 11.25  5 15 /
\put {$\s$, $s^r$} at 5.5 17
\put {\circle*{1.5}} [Bl] at 1 -1
\put {\circle*{1.5}} [Bl] at 2 0
\put {\circle*{1.5}} [Bl] at 3 3
\put {\circle*{1.5}} [Bl] at 4 8
\put {\circle*{1.5}} [Bl] at 5 15
\put {\scriptsize{${\bf\underline{4}}$}} [Bl] at 1.2 -2.5
\put {\scriptsize{${\bf\underline{9}}$}} [Bl] at 2.2 -1.5
\put {\scriptsize{${\bf\underline{16}}$}} [Bl] at 3 1
\put {\scriptsize{${\bf\underline{25}}$}} [Bl] at 4 6
\put {\scriptsize{${\bf\underline{36}}$}} [Bl] at 5 13
\plot 0 8  0.5 11.25  1 15  1.5 19.25  2 24 /
\put {$\t$, $t^r$} at 2.5 26
\put {\circle*{1.5}} [Bl] at 0 8
\put {\circle*{1.5}} [Bl] at 1 15
\put {\circle*{1.5}} [Bl] at 2 24
\put {\scriptsize{${\bf\underline{1}}$}} [Bl] at 0.2 6.5
\put {\scriptsize{${\bf\underline{4}}$}} [Bl] at 1 13
\put {\scriptsize{${\bf\underline{9}}$}} [Bl] at 2 22
\endpicture \]

The lowest mode for $\s_{I_1}$ is actually $k=2$, since the $k=1$
mode can be gauged away by residual shift transformations \cite{deg}.
This can also be seen from the
action \eqref{act}, where this mode is simply absent.
\subsection{Vector fields}
The equations of motion for the vector fields on $AdS_3$
are obtained from the (anti)self--duality equation for the three--form
field strength with indices $(\m\n a)$ and from
the $(\m a)$ component of the Einstein equation.

We start with the vector fields $Z_{\m}^{r\pm}$ transforming in
the fundamental representation of $SO(n)$.
The transverse part of \eqref{con1} is
\begin{equation}
\label{Zr}
P^{\mp}_{k+1}(Z^{r\pm})_{\m I_3}=-{\mc Q}_{2\mu I_3}^{r\pm}
\end{equation}
and it automatically implies that the vector fields $Z_{\m}^{r\pm}$
are transverse at the linearized order.
Clearly, the linearized equation for $Z_{\m}^r$ follows from the
action (\ref{act}).

Consider the vector fields being mixtures of the spin one components
of the graviton and the 2--form potential $B$. Their
equations of motion have their origin in the transverse parts of
the $(\m a)$ component of the Einstein equation and the $(\m\n a)$
component of the self--duality equation and we find
\begin{equation}
\label{h}
-\frac{1}{4}\text{Max}(h)_{\m I_3}^{\pm}+\frac{1}{4}(\D+1)h_{\m I_3}^{\pm}
-P^{\mp}_{k+1}(Z^{\pm})_{\m I_3} = {\mc Q}_{5\mu I_3}^{\pm},
\end{equation}
and
\begin{equation}
\label{Z}
P^{\pm}_{k+1}(Z^{\pm})_{\m I_3}+h_{\m I_3}^{\pm}=-{\mc Q}_{3\m I_3}^{\pm},
\end{equation}
where
\begin{equation}
\text{Max}(V)_{\m}\equiv\nabla_{\n}^2V_{\m}-\nabla^{\n}\nabla_{\m}V_{\n},
\end{equation}
which satisfies $(*\nabla)^2=\text{Max}$.

Equation \eqref{h} can be factorized as follows
\begin{equation}
\bigl(P^{\mp}_{k+1}\bigr)^{\n}_{\m}\bigl(P^{\pm}_{k+1}(h^{\pm})_{\n I_3}
+4Z^{\pm}_{\n I_3}\bigr)=-4{\mc Q}_{5\m I_3}^{\pm}.
\end{equation}
Therefore, one set of solutions is found by diagonalizing the
system of first order equations
\begin{align}
P^{\pm}_{k+1}(h^{\pm})_{\m I_3}+4Z^{\pm}_{\m I_3}&=0,\\
P^{\pm}_{k+1}(Z^{\pm})_{\m I_3}+h^{\pm}_{\m I_3}&=0.
\end{align}
A simple calculation shows that
\begin{equation}
A_{\m I_3}^{\pm}=\pm2Z_{\m I_3}^{\pm}-h_{\m I_3}^{\pm},\quad
C_{\m I_3}^{\pm}=\pm2Z_{\m I_3}^{\pm}+h_{\m I_3}^{\pm}
\end{equation}
satisfy the following equations
\begin{equation}
P^{\pm}_{k-1}(A^{\pm})_{\m I_3}=0,\quad P^{\pm}_{k+3}(C^{\pm})_{\m I_3}=0
\end{equation}
at the linearized order.

On the other hand, solving the linear equation \eqref{Z} and
substituting into \eqref{h}, introducing
the canonical fields $A_{\m}^{\pm}$ and $C_{\m}^{\pm}$ and using
$[P_m,P_{m'}]=0$, one obtains
\begin{align}
P^{\mp}_{k+1}P^{\pm}_{k-1}(A^{\pm})_{\m I_3} &=4{\mc
Q}_{5\m I_3}^{\pm}\mp2P^{\mp}_{k+1}({\mc Q^{\pm}})_{3\m I_3},\\
P^{\mp}_{k+1}P^{\pm}_{k+3}(C^{\pm})_{\m I_3} &=-4{\mc
Q}_{5\m I_3}^{\pm}\mp2P^{\mp}_{k+1}({\mc Q^{\pm}})_{3\m I_3}.
\end{align}
It is straightforward to write down an action that leads to the
second order equations for the fields $A_{\m}^{\pm}$ and
$C_{\m}^{\pm}$ but this is not what we need, since these fields
are related by the linear constraint \eqref{Z}, that in terms
of the canonical fields reads as
\begin{equation}
P^{\pm}_{k-1}(A^{\pm})_{\m I_3}+P^{\pm}_{k+3}(C^{\pm})_{\m I_3}
=\mp4{\mc Q}_{3\m I_3}.
\end{equation}
By using equation \eqref{Z}, \eqref{h} can be reduced to the form
\begin{equation}
\label{h2}\text{Max}(h)_{\m I_3}^{\pm}-(k-1)(k+3)h_{\m I_3}^{\pm}
-8(h_{\m I_3}^{\pm}\pm(k+1)Z_{\m I_3}^{\pm})
=4\bigl({\mc Q}_{3\m I_3}^{\pm}-{\mc Q}_{5\m I_3}^{\pm}\bigr).
\end{equation}
The equations \eqref{Z} and \eqref{h2} can be derived from the Lagrangian
\begin{equation}
{\mc L}^{\pm}=-\frac{1}{4}F_{\m\n}(h^{\pm})F^{\m\n}(h^{\pm})
-\frac{1}{2}(k-1)(k+3)h_{\m}^{\pm}h^{\m\pm}
-4\bigl(h_{\m}^{\pm}\pm(k+1)Z_{\m}^{\pm}\bigr)^2\mp4(k+1)
\e^{\m\n\r}Z_{\m}^{\pm}\partial_{\n}Z_{\r}^{\pm}.
\end{equation}
After substituting the canonical fields this Lagrangian turns
into the one given in section 2,
equation \eqref{vector}.

The results for the relevant vector modes ({\it c.f.} Table 1) are
summarized in Fig.\ 2:
\vskip 0.2cm
\[ \beginpicture \setquadratic
\setcoordinatesystem units <1cm,1.7mm>
\setplotarea x from 0 to 5, y from 0 to 27
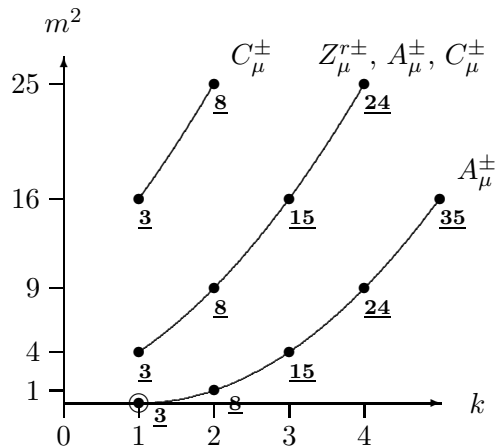
\axis bottom label {Fig.\ 2: Mass spectrum of vectors.}
ticks numbered from 0 to 4 by 1 /
\axis left ticks withvalues $1$ $4$ $9$ $16$ $25$ {} /
at 1 4 9 16 25 / /
\put {\vector(1,0){60}} [Bl] at 5 0
\put {$k$} at 5.5 0
\put {\vector(0,1){60}} [Bl] at 0 27
\put {$m^2$} at 0 30
\plot 1 4  2 9  3 16  3.5  20.25  4 25 /
\put {$Z_{\m}^{r\pm}$, $A_{\m}^{\pm}$, $C_{\m}^{\pm}$} at 4.5 27
\unitlength1mm
\put {\circle*{1.5}} [Bl] at 1 4
\put {\circle*{1.5}} [Bl] at 2 9
\put {\circle*{1.5}} [Bl] at 3 16
\put {\circle*{1.5}} [Bl] at 4 25
\put {\scriptsize{${\bf\underline{3}}$}} [Bl] at 1 2
\put {\scriptsize{${\bf\underline{8}}$}} [Bl] at 2 7
\put {\scriptsize{${\bf\underline{15}}$}} [Bl] at 3 14
\put {\scriptsize{${\bf\underline{24}}$}} [Bl] at 4 23
\plot 1 0  2 1  3 4  3.5 6.25  4 9  4.5 12.25  5 16 /
\put {$A_{\m}^{\pm}$} at 5.5 18
\put {\circle{2.5}}[Bl] at 1 0
\put {\circle*{1.5}} [Bl] at 1 0
\put {\circle*{1.5}} [Bl] at 2 1
\put {\circle*{1.5}} [Bl] at 3 4
\put {\circle*{1.5}} [Bl] at 4 9
\put {\circle*{1.5}} [Bl] at 5 16
\put {\scriptsize{${\bf\underline{3}}$}} [Bl] at 1.2 -1.5
\put {\scriptsize{${\bf\underline{8}}$}} [Bl] at 2.2 -0.5
\put {\scriptsize{${\bf\underline{15}}$}} [Bl] at 3 2
\put {\scriptsize{${\bf\underline{24}}$}} [Bl] at 4 7
\put {\scriptsize{${\bf\underline{35}}$}} [Bl] at 5 14
\plot 1 16  1.5 20.25  2 25 /
\put {$C_{\m}^{\pm}$} at 2.5 27
\put {\circle*{1.5}} [Bl] at 1 16
\put {\circle*{1.5}} [Bl] at 2 25
\put {\scriptsize{${\bf\underline{3}}$}} [Bl] at 1 14
\put {\scriptsize{${\bf\underline{8}}$}} [Bl] at 2 23
\endpicture \]

The Yang--Mills states $A^{I_3\pm}_{\m\,k=1}$ transform in the
adjoint representation of $SU(2)_L\times SU(2)_R$; they are pure
gauge and are components of the massless graviton multiplet,
which also contains
the non--propagating graviton and 4 gravitini.
\subsection{Symmetric tensor fields of second rank}
In principle, to find the equation of motion for the massive
gravitons \footnote{We loosely refer to symmetric
tensor fields coming from the $AdS_3$ components of
the metric as to massive gravitons.} on $AdS_3$
one has to consider the Einstein equation
\eqref{eqn} not only with indices $(\m\n)$, but also with
the indices $(\m a)$ and $(ab)$. The reason
for this is that the equations for
$\nabla^{\m}\varphi_{\m\n}$ and  $\varphi\equiv\varphi^{\m}_{\m}$
are constraints and do not follow from \eqref{eqn} if one considers only
the $(\m\n)$ components. Moreover,
the Einstein equation involves both the gravitons
and the scalar fields already at the linearized level and, therefore,
the procedure of constructing the quadratic Lagrangian becomes
highly non--trivial.
However, analogously to the case of type IIB supergravity
on $AdS_5\times S^5$ \cite{gleb7}, \cite{gleb1}, we can
replace the Einstein equation \eqref{eqn} by a new equivalent
equation from which the true equation for massive
gravitons follows from the $(\m\n)$ components.
``Equivalent'' means that the original and the new equations
coincide on--shell, i.e.,  when one takes into account the (anti)self--duality
equation for the field strength $H$.
To find the new equation we will use the shift of the graviton given in
equations \eqref{hmn}, \eqref{zeta} and \eqref{eta}.

At linear order we have\footnote{To simplify the notation throughout this section
we use $I\equiv I_1$.}
\begin{align}
R_{\m\n I}^{(1)}&=
R_{\m\n I}^{(1)}(\varphi)+\frac{8}{k+1}\N_{\m}\N_{\n}(\s_I-\t_I)
+4g_{\m\n}\left(\frac{k^2(k-1)}{k+1}\s_I
+\frac{(k+2)^2(k+3)}{k+1}\t_I\right)\nonumber\\
&+g_{\m\n}\left(-\frac{k(k-1)}{k+1}(\N_{\r}^2-m_{\s}^2)\s_I
+\frac{(k+2)(k+3)}{k+1}(\N_{\r}^2-m_{\t}^2)\t_I\right),
\end{align}
where
\begin{equation}
R_{\m\n I}^{(1)}(\varphi)\equiv-\frac{1}{2}\bigl((\N_{\r}^2-\D+6)
\varphi_{\m\n I}-\N_{\m}\N^{\r}\varphi_{\r\n I}-\N_{\n}\N^{\r}\varphi_{\r\m I}
+\N_{\m}\N_{\n}\varphi_I\bigr)+g_{\m\n}\varphi_I.
\end{equation}
Introducing the notation $H_MH_N\equiv H_{MPQ}{H_N}^{PQ}$ we obtain
\begin{align}
(H_{\m}H_{\n})^{(1)}_I&=2g_{\m\n}\varphi_I-2\varphi_{\m\n I}
+\frac{8}{k+1}\N_{\m}\N_{\n}(\s_I-\t_I)
+4g_{\m\n}\left(\frac{k^2(k-1)}{k+1}\s_I\right.
\nonumber\\
&\left.+\frac{(k+2)^2(k+3)}{k+1}\t_I\right)
+4g_{\m\n}\left(\frac{k-1}{k+1}(\N_{\r}^2-m_{\s}^2)\s_I
+\frac{k+3}{k+1}(\N_{\r}^2-m_{\t}^2)\t_I\right).
\end{align}
The first correction to the total curvature is found to be
\begin{align}
R^{(1)}_I&=
\N^{\m}\N^{\n}\varphi_{\m\n I}-\bigl(\N_{\m}^2-\D-2\bigr)\varphi_I\nonumber\\
&-2\left(\frac{k(k-1)}{k+1}(\N_{\m}^2-m_{\s}^2)\s_I
-\frac{(k+2)(k+3)}{k+1}(\N_{\m}^2-m_{\t}^2)\t_I\right).
\end{align}
Now we consider the combination
$\bigl(R_{\mu\nu}-\frac{1}{2}g_{\mu\nu}R-H_{\mu}H_{\nu}\bigr)^{(1)}_I$,
substitute the previous results and get
\begin{align}
\label{ref}\bigl(R_{\m\n}&-\frac{1}{2}g_{\m\n}R-H_{\m}H_{\n}\bigr)^{(1)}_I=
{\mc E}_{\m\n I}(\varphi)-2g_{\m\n}\varphi_I\nonumber\\
&-\frac{4}{k+1}g_{\m\n}\bigl((k-1)(\N_{\r}^2-m_{\s}^2)\s_I
+(k+3)(\N_{\r}^2-m_{\t}^2)\t_I\bigr).
\end{align}
Here
\begin{align}
{\mc E}_{\m\n}(\varphi)&\equiv
-\frac{1}{2}\bigl((\N_{\r}^2-\D+2)\varphi_{\m\n}-\N_{\m}\N^{\r}\varphi_{\r\n}
-\N_{\n}\N^{\r}\varphi_{\r\m}+g_{\m\n}\N^{\r}\N^{\l}\varphi_{\r\l}\nonumber\\
&+\N_{\m}\N_{\n}\varphi-g_{\m\n}(\N_{\r}^2-\D)\varphi\bigr).
\end{align}
Next we use
\begin{equation}
(H^2)^{(1)}_I=6\varphi_I+12\left(\frac{k-1}{k+1}(\N_{\m}^2-m_{\s}^2)\s_I
+\frac{k+3}{k+1}(\N_{\m}^2-m_{\t}^2)\t_I\right)
\end{equation}
and find that
\begin{equation}
\left(R_{\m\n}-\frac{1}{2}g_{\m\n}R-\bigl(H_{\m}H_{\n}
-\frac{1}{3}g_{\m\n}H^2\bigr)\right)^{(1)}={\mc E}_{\m\n}(\varphi)
\end{equation}
holds {\em off--shell}, i.e.\ without using the linearized equations
of motion for $\s_I$ and $\t_I$. Therefore the proper equation
for the massive gravitons is ({\it c.f.} \eqref{eqn})
\begin{equation}
\label{ein}
\begin{split}
R_{MN}-\frac{1}{2}g_{MN}R&=
H_M^IH_N^I-\frac{1}{3}g_{MN}H_{M_1M_2M_3}^IH^{I~M_1M_2M_3}\\
&+2P^{ir}_MP^{ir}_N-g_{MN}P^{ir}_LP^{ir~L}.
\end{split}
\end{equation}
At the linearized level the equation for massive gravitons is
${\mc E}_{\m\n I}(\varphi)\equiv {\mc E}_{\m\n}(\varphi^I_{\r\l})=0$.
{}For $k=0$ this  reduces to the known equation
for (massless) gravitons.
Note that the first line of \eqref{ein} {\em cannot} be obtained from
\begin{equation}
S\sim\int\sqrt{-g}\bigl(R-\frac{1}{3}H^2\bigr)
\end{equation}
since this would give a coefficient $1/6$ in front of $H^2$. However, recalling
that the field strengths are (anti)self--dual
we see that on--shell the term $H^2$ is zero and, therefore, its coefficient
in the Einstein equation remains arbitrary and may be fixed to any desired value.
Of course this just reflects the fact that
there is no simple covariant action when (anti)self--dual field strengths
are involved.

Taking the trace of ${\mc E}_{\m\n}(\varphi)$ one obtains
\begin{equation}
{\mc E}_{\m}^{\m}(\varphi)=-\frac{1}{2}\N_{\m}\bigl(\N_{\n}\varphi^{\m\n}
-\N^{\m}\varphi\bigr)-(\D+1)\varphi.
\end{equation}
On the other hand, the divergence of ${\mc E}_{\m\n}(\varphi)$ is
\begin{equation}
\N^{\m}{\mc E}_{\m\n}(\varphi)=\frac{\D}{2}\bigl(\N^{\m}\varphi_{\m\n}
-\N_{\n}\varphi\bigr).
\end{equation}
Thus for $k\neq0$ on--shell
massive gravitons are transverse and traceless at
the linear order.
\newpage
The result for the tensor modes ({\it c.f.} Table 1) is summarized in Fig.\ 3:
\vskip 0.2cm
\[ \beginpicture \setquadratic
\setcoordinatesystem units <1cm,1.7mm>
\setplotarea x from 0 to 5, y from 0 to 27
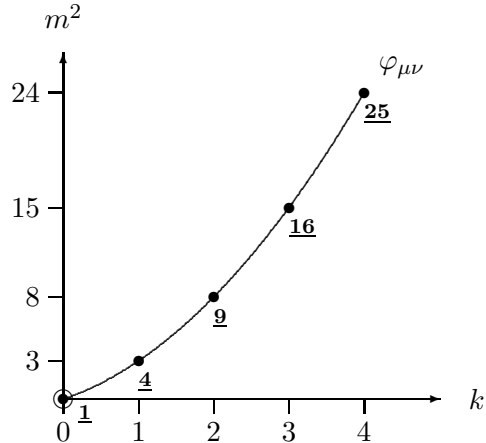
\axis bottom label {Fig.\ 3: Mass spectrum of symmetric tensors.}
ticks numbered from 0 to 4 by 1 /
\axis left ticks withvalues $3$ $8$ $15$ $24$ {} /
at 3 8 15 24 / /
\put {\vector(1,0){60}} [Bl] at 5 0
\put {$k$} at 5.5 0
\put {\vector(0,1){60}} [Bl] at 0 27
\put {$m^2$} at 0 30
\plot 0 0  1 3  2 8  3 15  4 24 /
\put {$\varphi_{\m\n}$} at 4.5 26
\unitlength1mm
\put {\circle*{1.5}} [Bl] at 0 0
\put {\circle{2.5}}[Bl] at 0 0
\put {\circle*{1.5}} [Bl] at 1 3
\put {\circle*{1.5}} [Bl] at 2 8
\put {\circle*{1.5}} [Bl] at 3 15
\put {\circle*{1.5}} [Bl] at 4 24
\put {\scriptsize{${\bf\underline{1}}$}} [Bl] at 0.2 -1.5
\put {\scriptsize{${\bf\underline{4}}$}} [Bl] at 1 1
\put {\scriptsize{${\bf\underline{9}}$}} [Bl] at 2 6
\put {\scriptsize{${\bf\underline{16}}$}} [Bl] at 3 13
\put {\scriptsize{${\bf\underline{25}}$}} [Bl] at 4 22
\endpicture \]
\section{Cubic couplings}
The cubic couplings involving at least two chiral primaries
were listed in section 2.1. In this section we
sketch their derivation\footnote{To simplify the
notation we drop the index $I$ that specifies a basis of an irreducible
representation
and denote $\s^{I_1}$ as $\s_1$ and similarly for the other fields.}.
As usual, the quadratic corrections involve higher derivative
terms and are in general non--Lagrangian. To make the couplings Lagrangian
and to remove the higher derivative terms one performs
nonlinear field redefinitions \cite{sei}.
To determine the cubic couplings it is sufficient to derive them from the
equation of motion of one of the fields involved.
However, the problem of finding 4--point functions requires to consider
all equations of motion. The reason is that the field redefinitions mentioned
above will induce contributions to quartic couplings \cite{gleb1}.
\subsection{Cubic couplings with scalar fields}
The cubic couplings of scalar fields corresponding to CPOs
were already obtained in \cite{mih}.
As an example we just consider the self--interaction of $\s$.
Keeping only the quadratic contributions one finds
the following structure\footnote{We do not present the explicit values
of the coefficients here and below because they are not very instructive.
They are explicitly given in \cite{ari}.}:
\begin{equation}\label{sigma}
\bigl(\N_{\m}^2-m_{\s}^2\bigr)\s_1=A_{123}\s_2\s_3
+B_{123}\N_{\m}\s_2\N^{\m}\s_3+C_{123}\N_{\m}\N_{\n}\s_2\N^{\m}\N^{\n}\s_3.
\end{equation}
To remove higher derivative terms one should make the field redefinition
\begin{equation}
\s_1\rightarrow \s_1+J_{123}\s_2\s_3+L_{123}\N_{\m}\s_2\N^{\m}\s_3,
\end{equation}
where
\begin{equation}
2L_{123}=C_{123},\qquad2J_{123}+L_{123}(m_2^2+m_3^2-m_1^2-4)=B_{123}.
\end{equation}
Then \eqref{sigma} takes the form
\begin{equation}
\bigl(\N_{\m}^2-m_{\s}^2\bigr)\s_1=
-\frac{3}{\kappa_1^{\s}}V^{\s\s\s}_{123}\s_2\s_3+\cdots,
\end{equation}
where $\cdots$ denote cubic terms induced by the field redefinitions,
$\kappa^{\s}=16k(k-1)$ and $V_{123}^{\s\s\s}$ as in \eqref{coupl}.
\subsection{Cubic couplings with vector fields}
We begin with the coupling $s^r\s Z_{\m}^{r\pm}$ and consider only
the quadratic corrections to the equations of motion for $Z^{r\pm}_{\m}$.
One finds ({\it c.f.} (\ref{Zr}))
\begin{equation}\label{Z1}
P^{\mp}_{m_Z}(Z^{r\pm})_{3\m}=A_{123}\N_{\mu}(\s_1s^r_2)
+B_{123}\s_1\N_{\m}s^r_2+C_{123}\N_{\m}\N_{\n}\s_1\N^{\n}s^r_2.
\end{equation}
To simplify the notation, here and below we suppress the label $\pm$
on the coefficients
describing the quadratic corrections.
After performing the field redefinition
\begin{equation}
Z^{r\pm}_{3\mu}\to Z^{r\pm}_{3\mu}\pm\N_{\mu}\L^r_3
+L_{123}P^{\pm}_{k_3+1}\bigl(\s_1\N s_2^r\bigr)_{\m},
\end{equation}
where
\begin{equation}
2L_{123}=-C_{123},\quad
(k_3+1)\L_3^r=(m_2^2L_{123}-A_{123})\s_1s^r_2+L_{123}\N_{\m}\s_1\N^{\m}s^r_2,
\end{equation}
equation \eqref{Z1} becomes
\begin{equation}
P^{\mp}_{m_Z}(Z^{r\pm})_{3\m}
=\frac{1}{8(k_3+1)}V^{\s sZ^{\pm}}_{123}\s_1\N_{\m}s^r_2.
\end{equation}

{}For the vector fields $A_{\m}^{\pm}$ and
$C_{\m}^{\pm}$ we restrict ourselves to the interaction
with $\s$; the case of $s^r$ is analogous.

Consider the quadratic corrections to the equations of motion
for $A_{\m}^{\pm}$. Keeping only terms quadratic in $\s$ we
find the following structure:
\begin{align}\label{Acorr}
P^{\mp}_{k_3+1}P^{\pm}_{k_3-1}(A^{\pm})_{3\m}&=
\N_{\m}\L_3+(V1)_{123}\N_{\m}\s_1\s_2+(V2)_{123}\N_{\n}\N_{\m}\s_1\N^{\n}\s_2+
(V3)_{123}\N_{\n}\N_{\l}\N_{\m}\s_1\N^{\n}\N^{\l}\s_2\nonumber\\
&\pm{\e_{\m}}^{\n\l}\N_{\n}\bigl((W0)_{123}\N_{\l}\s_1\s_2
+(W2)_{123}\N_{\r}\N_{\l}\s_1\N^{\r}\s_2\bigr),
\end{align}
where
\begin{equation}
\L_3=(\L0)_{123}\s_1\s_2+(\L2)_{123}\N_{\m}\s_1\N^{\m}\s_2.
\end{equation}
After the shift
\begin{equation}
A_{3\m}^{\pm}\to A_{3\m}^{\pm}+
\N_{\m}\bar{\L}_3+J_{123}\N_{\m}\s_1\s_2
+L_{123}\N_{\n}\N_{\m}\s_1\N^{\n}\s_2\pm
G_{123}{\e_{\m}}^{\n\l}\N_{\n}\bigl(\N_{\l}\s_1\s_2\bigr)
\end{equation}
with
\begin{equation}
\begin{split}
2L_{123}&=(V3)_{123},\quad 2G_{123}=(V3)_{123}+(W2)_{123},\\
2J_{123}&=(V2)_{123}+4G_{123}-L_{123}\bigl(m_1^2+m_2^2-6-(k+1)(k-1)\bigr),\\
-(k_3+1)(k_3-1)\bar{\L}_3&
=\L_3+\frac{1}{2}(m_1^2-m_2^2)\bigl((J_{123}-L_{123}-2G_{123})\s_1\s_2
+L_{123}\N_{\m}\s_1\N^{\m}\s_2\bigr)
\end{split}
\end{equation}
we can represent \eqref{Acorr} as
\begin{equation}\label{A}
\frac{1}{2}P^{\mp}_{k_3+1}P^{\pm}_{k_3-1}(A^{\pm})_{3\m}=
V_{123}^{\s\s A^{\pm}}\N_{\m}\s_1\s_2
\pm W_{123}^{\s\s A^{\pm}}P^{\pm}_{k_3-1}(\N\s_1\s_2)_{\m}.
\end{equation}
Analogously one finds for $C_{\m}^{\pm}$
\begin{equation}\label{C}
\frac{1}{2}P^{\mp}_{k_3+1}P^{\pm}_{k_3+3}(C^{\pm})_{3\m}=
V_{123}^{\s\s C^{\pm}}\N_{\m}\s_1\s_2
\pm W_{123}^{\s\s C^{\pm}}P^{\pm}_{k_3+3}(\N\s_1\s_2)_{\m}.
\end{equation}
The coefficients $W_{I_1I_2I_3}^{\s\s A^{\pm}}$
and $W_{I_1I_2I_3}^{\s\s C^{\pm}}$ are
\begin{align}
W_{I_1I_2I_3}^{\s\s A^{\pm}}&
=\frac{(\S+1)}{(k_1+1)(k_2+1)}\bigl((k_1-1)(k_1-k_3)
+(k_2-1)(k_2-k_3)\bigr)t_{I_1I_2I_3}^{\pm},\\
W_{I_1I_2I_3}^{\s\s C^{\pm}}&
=\frac{(2\a_3-1)}{(k_1+1)(k_2+1)}\bigl((k_1+1)(k_1+k_3)
+(k_2+1)(k_2+k_3)-4(k_3+1)\bigr)t_{I_1I_2I_3}^{\pm}.
\end{align}
{}From the equation of motion for $\s$ one gets
\begin{align}
\bigl(\N_{\m}^2-m_{\s}^2\bigr)\s_1=&
A^{A^{\pm}}_{123}\N^{\m}\s_2A_{3\m}^{\pm}
+B^{A^{\pm}}_{123}\N^{\m}\N^{\n}\s_2\N_{\m}A_{3\n}^{\pm}+
A^{C^{\pm}}_{123}\N^{\m}\s_2C_{3\m}^{\pm}
+B^{C^{\pm}}_{123}\N^{\m}\N^{\n}\s_2\N_{\m}C_{3\n}^{\pm}
\nonumber\\
&\pm D_{123}^{\pm}\N^{\m}\s_2
\bigl(P^{\pm}_{k_3+3}(C^{\pm})_{3\m}-P^{\pm}_{k_3-1}(A^{\pm})_{3\m}\bigr),
\end{align}
where $A^{A^{\pm}}_{123}$, $B^{A^{\pm}}_{123}$, $A^{C^{\pm}}_{123}$,
$B^{C^{\pm}}_{123}$ and $D^{\pm}_{123}$ are known coefficients \cite{ari}.

Performing the field redefinition
\begin{equation}
\s_1\rightarrow\s_1+\frac{1}{2}B^{A\pm}_{123}\N^{\m}\s_2A_{3\m}^{\pm}
+\frac{1}{2}B^{C\pm}_{123}\N^{\m}\s_2C_{3\m}^{\pm}
\end{equation}
we represent the result in the following form:
\begin{align}\label{AC}
\bigl(\N_{\m}^2-m_{\s}^2\bigr)\s_1=&-\frac{2}{\kappa_1^{\s}}
\left(V^{\s\s A^{\pm}}_{123}\N^{\m}\s_2A_{3\m}^{\pm}
+V^{\s\s C^{\pm}}_{123}\N^{\m}\s_2C_{3\m}^{\pm}\right.\nonumber\\
&\left.\pm\N^{\m}\s_2\bigl((W^{\s\s A^{\pm}}_{123}
+\a_{123}^{\pm})P^{\pm}_{k_3-1}(A^{\pm})_{3\m}+(W^{\s\s C^{\pm}}_{123}
+\a_{123}^{\pm})P^{\pm}_{k_3+3}(C^{\pm})_{3\m}\bigr)\right)\nonumber\\
&\pm\bigl((d\Omega)_{123}+\frac{2}{\kappa_1^{\s}}\a_{123}^{\pm}\bigr)
\N^{\m}\s_2\bigl(P^{\pm}_{k_3-1}(A^{\pm})_{3\m}
+P^{\pm}_{k_3+3}(C^{\pm})_{3\m}\bigr).
\end{align}
Here
\begin{equation}
B^{A^{\pm}}_{123}+D^{\pm}_{123}=
\frac{2}{\kappa_1^{\s}}W^{\s\s A^{\pm}}_{123}-(d\Omega)_{123},\qquad
B^{C^{\pm}}_{123}+D^{\pm}_{123}=
-\frac{2}{\kappa_1^{\s}}W^{\s\s C^{\pm}}_{123}+(d\Omega)_{123}
\end{equation}
and $\a_{123}^{\pm}$ are yet to be determined $SO(4)$ tensors
that are {\em antisymmetric} in 1 and 2. The last line in \eqref{AC}
is proportional to the linear constraint (\ref{cons})
and, therefore, only contributes to the next order.
The tensors $\a^{\pm}_{123}$ are fixed as follows.
Equation \eqref{AC} can be derived from
\begin{align}
{\mc L}_3^{\s}(A_{\m}^{\pm},C_{\m}^{\pm})&=
V^{\s\s A^{\pm}}_{I_1I_2I_3}\s_{I_1}\N^{\m}\s_{I_2}A_{\m I_3}^{\pm}+
V^{\s\s C^{\pm}}_{I_1I_2I_3}\s_{I_1}\N^{\m}\s_{I_2}C_{\m I_3}^{\pm}\\
&\pm\bigl(W^{\s\s A^{\pm}}_{I_1I_2I_3}+\a_{I_1I_2I_3}^{\pm}\bigr)
P^{\pm}_{k_3-1}(\s_{I_1}\N\s_{I_2})^{\m}A_{\m I_3}^{\pm}
\pm\bigl(W^{\s\s C^{\pm}}_{I_1I_2I_3}+\a_{I_1I_2I_3}^{\pm}\bigr)
P^{\pm}_{k_3+3}(\s_{I_1}\N\s_{I_2})^{\m}C_{\m I_3}^{\pm}.\nonumber
\end{align}

Varying this Lagrangian with respect to $A_{\m}^{\pm}$ and
$C_{\m}^{\pm}$ we can write down the quadratic corrections to the equations
of motion for the vector fields and compare them with our
results \eqref{A} and \eqref{C}. Then we find that
\begin{equation}
\a_{I_1I_2I_3}^{\pm}=-\frac{1}{2}\bigl(W^{\s\s A^{\pm}}_{I_1I_2I_3}
+W^{\s\s C^{\pm}}_{I_1I_2I_3}\bigr).
\end{equation}
Substituting the shifts of $A_{\m}$ and $C_{\m}$ we finally
represent \eqref{AC} as
\begin{align}
\kappa_1^{\s}\bigl(\N_{\m}^2-m_{\s}^2\bigr)\s_1=&
-V^{\s\s A^{\pm}}_{123}\bigl(2\N^{\m}\s_2A_{3\m}^{\pm}
+\s_2\N^{\m}A_{3\m}^{\pm}\bigr)
\mp W^{\s\pm}_{123}\bigl(2\N^{\m}\s_2P^{\pm}_{k_3-1}
(A^{\pm})_{3\m}\pm(k_3-1)\s_2\N^{\m}A_{3\m}^{\pm}\bigr)
\nonumber\\
&-V^{\s\s C^{\pm}}_{123}\bigl(2\N^{\m}\s_2C_{3\m}^{\pm}
+\s_2\N^{\m}C_{3\m}^{\pm}\bigr)
\pm W^{\s\pm}_{123}\bigl(2\N^{\m}\s_2P^{\pm}_{k_3+3}
(C^{\pm})_{3\m}\pm(k_3+3)\s_2\N^{\m}C_{3\m}^{\pm}\bigr)
\nonumber\\
&+\text{{\em cubic}},
\end{align}
where
\begin{equation}
W_{123}^{\s\pm}=\frac{1}{2}\bigl(W^{\s\s A^{\pm}}_{123}
-W^{\s\s C^{\pm}}_{123}\bigr).
\end{equation}
Note that the terms on the r.h.s. quadratic in fields follow
from action (\ref{act}),
while ``cubic'' denotes the cubic terms relevant
only for determining the quartic couplings.
\subsection{Cubic couplings with massive gravitons}
Here the most complicated part is to derive the interaction of
$\sigma$ with massive gravitons from the equation of motion for
$\varphi_{\m\n}$, because $\sigma$ is constructed by using the metric itself.
Since the analysis of the interaction of $\varphi_{\m\n}$ with $s^r$
(although much simpler) proceeds along the same lines we will omit it.
After careful computation one finds the following structure:
\begin{align}\label{varphi}
{\mc E}_{1\m\n}(\varphi)&=
(A2)_{123}\N_{\m}\s_2\N_{\n}\s_3+(A4)_{123}\N_{\r}\N_{\m}\s_2\N^{\r}\N_{\n}\s_3
+(A6)_{123}\N_{\r}\N_{\l}\N_{\m}\s_2\N^{\r}\N^{\l}\N_{\n}\s_3\nonumber\\
&+(B2)_{123}\bigl(\N_{\m}(\s_2\N_{\n}\s_3)+\N_{\n}(\s_2\N_{\m}\s_3)\bigr)
+(B4)_{123}\bigl(\N_{\m}(\N_{\r}\s_2\N^{\r}\N_{\n}\s_3)\nonumber\\
&+\N_{\n}(\N_{\r}\s_2\N^{\r}\N_{\m}\s_3)\bigr)+g_{\m\n}C_1,
\end{align}
where
\begin{equation}
C_1 =(C0)_{123}\s_2\s_3+(C2)_{123}\N_{\m}\s_2\N^{\m}\s_3
+(C4)_{123}\N_{\m}\N_{\n}\s_2\N^{\m}\N^{\n}\s_3
-\frac{1}{2}(A6)_{123}\N_{\m}\N_{\n}\N_{\r}\s_2\N^{\m}\N^{\n}\N^{\r}\s_3.
\end{equation}
Performing the shift of the graviton
\begin{equation}\label{gsh}
\varphi_{1\m\n}={\varphi'}_{1\m\n}+\N_{\m}\xi_{1\n}+\N_{\n}\xi_{1\m}
+g_{\m\n}\eta_1+K_{123}\N_{\m}\s_2\N_{\n}\s_3
+L_{123}\N_{\r}\N_{\m}\s_2\N^{\r}\N_{\n}\s_3,
\end{equation}
where
\begin{equation}
\label{xi}
\begin{split}
L_{123} &=-(A6)_{123},\quad
K_{123}=-(A4)_{123}+\frac{1}{2}(A6)_{123}(m_2^2+m_3^2-\D_1-10),\\
\frac{\Delta_1}{2}\xi_{1\m}&=\frac{1}{4}\N_{\m}\eta_1+M_{123}\s_2\N_{\m}\s_3
+N_{123}\N_{\n}\s_2\N^{\n}\N_{\m}\s_3
\end{split}
\end{equation}
and $M_{123}$, $N_{123}$ are given by
\begin{equation}
M_{123}=(B2)_{123}-\frac{3}{2}m_2^2(A6)_{123}-\frac{1}{2}m_2^2K_{123},\quad
N_{123}=(B4)_{123}+\frac{1}{2}(m_2^2-1)(A6)_{123}
\end{equation}
equation \eqref{varphi} takes the form
\begin{equation}\label{varphi1}
{\mc E}_{1\m\n}(\varphi')=V^{\s\s\varphi}_{123}\N_{\m}\s_2\N_{\n}\s_3
+g_{\m\n}\bar{C}_1.
\end{equation}
Here
\begin{equation}
\bar{C}_1=(\bar{C}0)_{123}\s_2\s_3+(\bar{C}2)_{123}\N_{\m}\s_2\N^{\m}\s_3
+(\D_1+1)\eta_1.
\end{equation}

Thus only $\eta$ has not been fixed yet. In fact, a change of $\eta$
(with a simultaneous change of $\xi$ according
to \eqref{xi}) amounts only to a change of the interaction of the trace
of the massive gravitons with the chiral primaries. In particular,
it is possible to choose $\eta$ in such a way that only the traceless
part of $\varphi_{\m\n}$ interacts with $\sigma$.
But as was pointed out in \cite{gleb1}, for the case of $AdS_5$ this
choice leads to the appearance of quartic couplings with
six derivatives, which are absent only if $\eta$ is chosen such that
${\varphi'}_1=(T0)_{123}\sigma_2\sigma_3$. We expect that
this will also be the case
for $AdS_3$ and therefore follow this approach.

Taking the trace and divergence of \eqref{varphi1} and representing
\begin{equation}
\eta_1=A_{123}\s_2\s_3+B_{123}\N_{\m}\s_2\N^{\m}\s_3,
\end{equation}
where
\begin{equation}
(\D_1+1)A_{123}=-(\bar{C}0)_{123}-\frac{1}{4}(m_2^2+m_3^2-\D_1)
V^{\s\s\varphi}_{123},\quad
(\D_1+1)B_{123}=-(\bar{C}2)_{123}-\frac{1}{2}V^{\s\s\varphi}_{123},
\end{equation}
we find that
\begin{equation}
{\varphi'}_1=\frac{V^{\s\s\varphi}_{123}}{4\D_1(\D_1+1)}
\bigl((m_2^2-m_3^2)^2+2\D_1(m_2^2-m_3^2)-3\D_1^2\bigr)\s_2\s_3.
\end{equation}
Then the final result for the interaction of $\varphi_{\m\n}$
with $\s$ has the nice form
\begin{equation}
{\mc E}_{1\m\n}(\varphi')=V^{\s\s\varphi}_{123}\bigl(\N_{\m}\s_2\N_{\n}\s_3
-\frac{1}{2}g_{\m\n}(\N_{\r}\s_2\N^{\r}\s_3
+\frac{1}{2}(m_2^2+m_3^2-\D_1)\s_2\s_3)\bigr),
\end{equation}
which is the natural generalization of the interaction of the
massless graviton with scalar fields.

On the other hand from the equation of motion for $\s$ one obtains
\begin{equation}
\bigl(\N_{\m}^2-m_{\s}^2\bigr)\s_1
=\frac{2}{\kappa_1^{\s}}V^{\s\s\varphi}_{123}\N^{\m}\N^{\n}\s_2\varphi_{3\m\n}
+K_{123}^{\varphi}\N^{\m}\s_2\N^{\n}\varphi_{3\m\n}
+T_{123}^{\varphi}\s_2\varphi_3.
\end{equation}

Substituting the shift \eqref{gsh} of the graviton we represent the equation as
\begin{equation}
\kappa_1^{\s}\bigl(\N_{\m}^2-m_{\s}^2\bigr)\s_1=
V^{\s\s\varphi}_{123}\left(2\N^{\m}\bigl(\N^{\n}\s_2{\varphi'}_{3\m\n}\bigr)
-\N^{\m}\bigl(\N_{\m}\s_2{\varphi'}_3\bigr)
+\frac{1}{2}(m_1^2+m_2^2-\D_3)\s_2{\varphi'}_3\right)+\text{{\em cubic}}.
\end{equation}
Again the terms on the r.h.s. quadratic in fields follow from
action (\ref{act}), while ``cubic'' again denotes
the cubic terms relevant only for determining the quartic couplings.
\vskip 0.2cm
\noindent{\large\bf Acknowledgements}\\
G. A. is grateful to S. Frolov for many valuable discussions and to
A. Jevicki for helpful
correspondence. We wish to acknowledge the
hospitality of the E.S.I. in Vienna where part of the work was performed.
This work is supported by GIF -- the German-Israeli Foundation for
Scientific Research and by the TMR programme ERBRMX-CT96-0045.
G. A. was supported by the Alexander von Humboldt Foundation and in part
by the RFBI grant N99-01-00166 and by INTAS-99-1782.
\section*{Appendix A}
\setcounter{section}{1}
\renewcommand{\thesection}{\Alph{section}}
\setcounter{equation}{0}
Here we give the results for the expansion of the Einstein
equation \eqref{eqn}, the equations of motion for the scalar
fields \eqref{scalar} and the (anti)self--duality equation
\eqref{sat} up to the second order both in spacetime and
coset metric perturbations. The various quantities defined
in \eqref{cart} and \eqref{fs} are given by:
\begin{align}
Q^{ij}_M&=\frac{1}{2}(\phi^{ir}\N_M\phi^{jr}-\N_M\phi^{ir}\phi^{jr}),\\
Q^{rs}_M&=\frac{1}{2}(\phi^{jr}\N_M\phi^{js}-\N_M\phi^{jr}\phi^{js}),\\
P^{ir}_M&=\frac{1}{\sqrt{2}}\N_M\phi^{ir},\\
\label{Hi}
H^i&=\d^{i5}\e+g^i+\phi^{ir}g^r+\frac{1}{2}\e\phi^{ir}\phi^{r5},\\
\label{Hr}
H^r&=g^r+\e\phi^{5r}+g^i\phi^{ir}.
\end{align}

Decomposing the Einstein equation \eqref{eqn} up to the
second order in $h_{MN}$ one finds
\begin{align}
R_{MN}^{(1)}+R_{MN}^{(2)}&=
H^I_{MPQ}{H^I_N}^{PQ}-2\bigl(h^{KL}-{h^2}^{KL}\bigr)H^I_{MKQ}{H^I_{NL}}^Q
\nonumber\\
&+h^{K_1L_1}h^{K_2L_2}H^I_{MK_1K_2}H^I_{NL_1L_2}+\N_M\phi^{ir}\N_N\phi^{ir}.
\end{align}
Here
\begin{align}
{R^{(1)}}_{MN}&=\N_Kh^K_{MN}-\frac{1}{2}\N_M\N_Nh,\\
{R^{(2)}}_{MN}&=-\N_K\left(h^K_Lh^L_{MN}\right)+\frac{1}{4}\N_M\N_N
\left(h^{KL}h_{KL}\right)+\frac{1}{2}h^K_{MN}\N_Kh-h^K_{ML}h^L_{KN},
\end{align}
where
\begin{equation}
h^K_{MN}\equiv\frac{1}{2}\left(\N_Mh^K_N+\N_Nh^K_M-\N^Kh_{MN}\right),
\qquad h\equiv h^K_K.
\end{equation}

Decomposing the equations of motion for the scalar fields $\phi^{ir}$
\eqref{scalar} up to the second order we obtain
\begin{equation}
\bigl(\N_{\m}^2+\N_a^2\bigr)\phi^{ir}=
\N_K\phi^{ir}\left(\N_Lh^{KL}-\frac{1}{2}\N^Kh\right)+h^{KL}\N_K\N_L\phi^{ir}
+\frac{2}{3}H_{KLM}^i\left({H^r}^{KLM}-3h^K_S{H^r}^{SLM}\right).
\end{equation}

Finally, the (anti)self--duality equation \eqref{sat}, expanded to
second order in $h_{MN}$ is:
\begin{equation}
H\mp*H\pm T^{(1)} \pm   T^{(2)}=0,
\end{equation}
where we have introduced the following notation:
\begin{align}
\label{T1}
T^{(1)}_{M_1M_2M_3}&\equiv
\frac{1}{2}h(*H)_{M_1M_2M_3}-3\,h^K_{[M_1}(*H)_{M_2M_3]K},\\
\label{T2}
T^{(2)}_{M_1M_2M_3}&\equiv
\frac{3}{2}hh^K_{[M_1}(*H)_{M_2M_3]K}-\left(\frac{1}{8}h^2
+\frac{1}{4}h^{KL}h_{KL}\right)(*H)_{M_1M_2M_3}
-3h^K_{[M_1}h^L_{M_2}(*H)_{M_3]KL}.
\end{align}
Recall that the operation $*$ is with respect to the background metric.
Equations \eqref{T1} and \eqref{T2} have to be combined with
\eqref{Hi} and \eqref{Hr}.
\section*{Appendix B}
\setcounter{section}{2}
\renewcommand{\thesection}{\Alph{section}}
\setcounter{equation}{0}
Following \cite{sei} we establish the formulae for integrals
of spherical harmonics on $S^3$ used in the paper.

In deriving the equations of motion for the various scalar fields
we encountered a number of integrals of scalar spherical
harmonics, all of which can be reduced to $a_{I_1I_2I_3}$ ({\it c.f.} \eqref{notation2}).
In terms of $\D_i\equiv k_i(k_i+2)$ we find:
\begin{equation}
\begin{split}
\int_{S^3}\N^aY^{I_1}\N_aY^{I_2}&=\D_1\d^{I_1I_2},\\
\int_{S^3}\N^{(a}\N^{b)}Y^{I_1}\N_a\N_bY^{I_2}&
=\frac{2}{3}\D_1(\D_1-3)\d^{I_1I_2}
\end{split}
\end{equation}
and
\begin{align}
\int_{S^3}\N^aY^{I_1}\N_aY^{I_2}Y^{I_3}&
=\frac{1}{2}(\D_1+\D_2-\D_3)a_{I_1I_2I_3},\nonumber\\
\int_{S^3}\N^{(a}\N^{b)}Y^{I_1}\N_a\N_bY^{I_2}Y^{I_3}&=
\left(\frac{1}{4}(\D_1+\D_2-\D_3)(\D_1+\D_2-\D_3-4)
-\frac{1}{3}\D_1\D_2\right)a_{I_1I_2I_3},\\
\int_{S^3}\N_{(a}Y^{I_1}\N_{b)}Y^{I_2}\N^a\N^bY^{I_3}&=
\left(\frac{1}{4}(\D_1+\D_3-\D_2)(\D_2+\D_3-\D_1)
+\frac{1}{6}\D_3(\D_1+\D_2-\D_3)\right)a_{I_1I_2I_3}.\nonumber
\end{align}
In the computation of interaction vertices involving vector fields
one needs the following integrals:
\begin{equation}
\begin{split}
\int_{S^3}\N^{(a}\N^{b)}Y^{I_1}Y^{I_2}\N_aY^{I_3\pm}_b&
=\frac{1}{2}(\D_1-\D_2+\D_3-3)t_{I_1I_2I_3}^{\pm},\\
\int_{S^3}\N^{(a}\N^{b)}Y^{I_1}\N_aY^{I_2}Y^{I_3\pm}_b&
=\frac{1}{2}\left(\frac{1}{3}\D_1+\D_2-\D_3\right)t_{I_1I_2I_3}^{\pm}.
\end{split}
\end{equation}
Finally, for tensor spherical harmonics $Y_{(ab)}^{I_5\pm}$ we used
\begin{equation}
\begin{split}
\int_{S^3}\N^{(a}\N^{b)}Y^{I_1}\N_a\bigl(\N^cY^{I_2}Y^{I_3\pm}_{(bc)}\bigr)
=\frac{2}{3}(\D_1-3)p_{I_1I_2I_3}^{\pm},\\
\int_{S^3}\N^a\N^bY^{I_1}\N^cY^{I_2}\N_cY^{I_3\pm}_{(ab)}
=\frac{1}{2}(\D_1-\D_2-\D_3-4)p_{I_1I_2I_3}^{_\pm}.
\end{split}
\end{equation}

\end{document}